\documentclass[fleqn,usenatbib]{mnras}

\usepackage{newtxtext,newtxmath}
\usepackage[T1]{fontenc}
\usepackage{ae,aecompl}

\DeclareRobustCommand{\VAN}[3]{#2}
\let\VANthebibliography\thebibliography
\def\thebibliography{\DeclareRobustCommand{\VAN}[3]{##3}\VANthebibliography}

\usepackage{graphicx}	% Including figure files
\usepackage{amsmath}	% Advanced maths commands
\usepackage{amssymb}	% Extra maths symbols
\usepackage{gensymb}
\usepackage[export]{adjustbox}
\usepackage{longtable}

\title[Runaway and walkaway stars from the ONC]{Runaway and walkaway stars from the ONC with \textit{Gaia} DR2}

\author[Schoettler et al.]{Christina Schoettler,$^{1,2}$\thanks{E-mail: cschoettler1@sheffield.ac.uk}
Jos de Bruijne,$^{2}$
Eero Vaher$^{2,3}$
and Richard J. Parker$^{1}$\thanks{Royal Society Dorothy Hodgkin Fellow}
\\
$^{1}$Department of Physics and Astronomy, The University of Sheffield, Hicks Building, Hounsfield Road, Sheffield S3 7RH, UK\\
$^{2}$Science Support Office, Directorate of Science, European Space Research and Technology Centre (ESA/ESTEC)\\
Keplerlaan 1, NL-2201 AZ Noordwijk, The Netherlands\\
$^{3}$Lund Observatory, Department of Astronomy and Theoretical Physics,
Lund University, Box 43, SE-221 00 Lund, Sweden}

\date{Accepted XXX. Received YYY; in original form ZZZ}

\pubyear{2020}

\begin{document}
\label{firstpage}
\pagerange{\pageref{firstpage}--\pageref{lastpage}}
\maketitle

\begin{abstract}
Theory predicts that we should find fast, ejected (runaway) stars of all masses around dense, young star-forming regions. $N$-body simulations show that the number and distribution of these ejected stars could be used to constrain the initial spatial and kinematic substructure of the regions. We search for runaway and slower walkaway stars within 100 pc of the Orion Nebula Cluster (ONC) using \textit{Gaia} DR2 astrometry and photometry. We compare our findings to predictions for the number and velocity distributions of runaway stars from simulations that we run for 4 Myr with initial conditions tailored to the ONC. In \textit{Gaia} DR2, we find 31 runaway and 54 walkaway candidates based on proper motion, but not all of these are viable candidates in three dimensions. About 40 per cent are missing radial velocities, but we can trace back 9 3D-runaways and 24 3D-walkaways to the ONC, all of which are low/intermediate-mass (<8 M$_{\sun}$). Our simulations show that the number of runaways within 100 pc decreases the older a region is (as they quickly travel beyond this boundary), whereas the number of walkaways increases up to 3 Myr. We find fewer walkaways in \textit{Gaia} DR2 than the maximum suggested from our simulations, which may be due to observational incompleteness. However, the number of \textit{Gaia} DR2 runaways agrees with the number from our simulations during an age of $\sim$1.3--2.4 Myr, allowing us to confirm existing age estimates for the ONC (and potentially other star-forming regions) using runaway stars.
\end{abstract}

\begin{keywords}
astrometry -- stars: kinematics and dynamics -- open clusters and associations: individual: Orion Nebula Cluster 
\end{keywords}

%%%%%%%%%%%%%%%%%%%%%%%%%%%%%%%%%%%%%%%%%%%%%%%%%%

%%%%%%%%%%%%%%%%% BODY OF PAPER %%%%%%%%%%%%%%%%%%c

%\begingroup
%\let\clearpage\relax
%\tableofcontents
%\endgroup

\section{Introduction}

Stars often form in grouped or clustered environments, i.e. star-forming regions where we observe higher stellar densities than in the Galactic field \citep{RN25, RN59}. These young star-forming regions undergo dynamical evolution that can rapidly change their spatial and kinematic structure in only a few Myr \citep[e.g.][]{RN4, RN258}. This dynamical evolution can lead to the erasure of initial substructure \citep[e.g.][]{RN14,RN4,RN248}, a reduction in stellar density \citep[e.g.][]{RN277,RN8}, the destruction of primordial binaries/multiples \citep[e.g.][]{RN261,RN256,RN277,RN257} and dynamical mass segregation \citep[e.g][]{2007ApJ...655L..45M,RN15,2009MNRAS.396.1864M,2009MNRAS.400..657M,RN5}. It can also affect young planetary systems and protoplanetary discs around young stars in these regions, as they can be disrupted during a dense phase \citep[e.g.][]{RN271,RN272,RN269,2016ApJ...828...48V,2019MNRAS.485.4893N}. While these regions dynamically evolve, they also eject a proportion of their member stars \citep[e.g.][]{RN312,RN309} at velocities that can reach $\sim$400\,km\,s$^{-1}$ \citep[e.g.][]{RN241}.

Fast, ejected, massive stars were first discovered over 60 years ago \citep{RN325, RN255} and termed runaway stars \citep[RW,][]{RN67}. Classically, ejected stars are considered to fall into the runaway velocity regime when their peculiar velocity exceeds $\sim$30-40\,km\,s$^{-1}$ or when they are found far away from star-forming regions or the Galactic plane \citep[e.g.][]{RN255, RN67, RN276, RN50, RN190, RN137,2011A&A...530L..14B,RN293}. The velocity boundary has recently been proposed to be relaxed to include slower ejected stars as these can also travel several tens of pc within the lifetime of star-forming regions, resulting in stars being found far outside their natal regions. \citet{RN137} suggested a lower limit of $\sim$5\,km\,s$^{-1}$ for these slow runaway stars and \citet{RN136} subsequently termed them walkaway stars (WW).

RW and WW stars can both be created via one of two formation mechanisms. The first is the binary supernova scenario \citep{RN67}, which happens when the more massive primary star in a close binary undergoes a core-collapse supernova. The secondary, less massive companion star is then ejected with a velocity which is a large fraction of its previous orbital velocity as a consequence of the binary being completely disrupted. The second mechanism is the dynamical ejection scenario \citep{RN189}, which does not involve supernovae and can therefore explain ejections from very young star-forming regions. These ejections are the result of dynamical interactions where close encounters between at least a single, massive star with a binary, or a binary-binary interaction can lead to ejection of one or more of these stars.

Most of the currently known RW stars are high-mass (OB) stars \citep[see][]{RN263}. This is partly due to their intrinsic high luminosity, making these massive stars easier to observe compared to lower-mass stars. Only a few low-mass RW star detections have been made up to now \citep{RN252, 2019JKAS...52...57Y}. \citet{RN309} used $N$-body simulations to show that young star-forming regions eject RW and WW stars across a wide mass range (0.1--50 M$_{\sun}$). They suggested that we should be able to find a number of low/intermediate-mass RWs/WWs in addition to massive ones around many nearby, young star-forming regions and that the number and velocity distributions of RW/WW stars could allow us to constrain the initial conditions of star formation and the dynamical histories of these regions.

The limitations (observational bias) in searching for lower-mass RW and WW stars have reduced considerably thanks to the second \textit{Gaia} data release in April 2018 \citep[DR2,][]{RN308, RN238}. The astrometry of \textit{Gaia} DR2 has been instrumental in the continuing search for RW and the even faster hypervelocity stars \citep[e.g.][]{RN316,RN320,RN252,RN315,RN313,RN319,RN323,2018ApJ...866...39B,2020MNRAS.tmp...34R}. 

\textit{Gaia} DR2 \citep{RN308, RN238} provides five-parameter astrometry with position, parallax and proper motion information for over 1.3 billion sources down to an apparent G-magnitude limit of $G \approx$ 21 mag. This allows us to probe down to the low-mass end of the initial mass function (IMF) for potential RW and WW stars from nearby star-forming regions (<1 kpc). In addition to highly accurate astrometry, photometry information for roughly the same number of sources is available in three passbands ($G$, $G_{\rm{BP}}$ and $G_{\rm{RP}}$) allowing us to construct colour magnitude diagrams (CMD) of potential RW and WW candidates and deduce ages using PARSEC isochrones \citep{RN225}. 

In this paper, we use $N$-body simulations to predict the number of RW and WW stars for an ONC-like star-forming region. We then use \textit{Gaia} DR2 observations to search for RW and WW stars around the ONC. This paper is organised as follows. In Section 2, we describe our target and the \textit{Gaia} DR2 data selection process. Section 3 describes the \textit{Gaia} DR2 data analysis process. This is followed in Section 4 by a brief description of the simulation set-up and the predictions from our simulations. The observational results are presented in Section 5, followed by a brief discussion in Section 6. Section 7 provides concluding remarks for this paper. 

\section{Target and data selection}

\subsection{Search target}
Our target for the search for RW and WW stars is the ONC. This star-forming region is well-suited for this purpose due to its proximity to Earth \citep[$\sim$400 pc, e.g.][]{RN333,RN264}. At such a close distance, the faintest stars in our data set will be stars down to sub-solar masses with reasonably accurate \textit{Gaia} DR2 proper motions \citep{RN264}. The ONC is a very young region and its ejected stars will be easier to trace back as they are more likely to be still be in close proximity to the region. It has a mean age estimate of 2--3 Myr and a spread of $\sim$2--2.5 Myr \citep{RN218,RN212}. For our analysis, we adopt an upper age limit of 4 Myr, thus encompassing the full range of estimated ages of the ONC.

The ONC has an established list of cluster members across different wavelength bands, which allows us to define a clear cluster boundary to trace back ejected stars. \citet{1997AJ....113.1733H} and \citet{1998ApJ...492..540H} suggested that the number of stars visible in the optical spectrum is $\sim$1600 and a further $\sim$1900 stars are visible in the infrared (IR). An updated census by \citet{2012ApJ...748...14D} increased the number of known members in the optical spectrum to $\sim$1750 stars. This large number of cluster stars and higher local stellar density can increase the likelihood of dynamical ejections \citep[e.g.][]{RN241, RN312, RN309}. \citet{1997AJ....113.1733H} considered the projected size of the ONC in two dimensions to be 2.5 $\times$ 4.5 pc and \citet{RN321} suggested the cluster to have a nominal radius of $\sim$2.5 pc. The ONC members in the \citet{2012ApJ...748...14D} census also do not extend any further than this radius on the sky. For our analysis, we therefore use this nominal radius as our cluster boundary on the sky. We consider the ONC to be the region associated with the nebula centred around the Trapezium cluster.

To become a RW/WW, a star should be unbound from its birth region meaning it has to at least reach the respective escape velocity of the region. \citet{RN322} suggested an upper limit on the angular escape speed from the ONC of $\sim$\,3.1\,mas\,yr$^{-1}$. At an adopted distance of 400 pc \citep[e.g.][]{RN333,RN264} to the region, this angular escape speed translates to a space velocity of $\sim$5.8\,km\,s$^{-1}$. This implies that the suggested lower velocity limit for walkaways of 5\,km\,s$^{-1}$ \citep{RN137} might not be appropriate for star-forming regions with a total mass similar to or higher than the ONC. This lower velocity limit value can then result in considering stars to have ``walked away'' while still being gravitationally bound to their birth region. For our analysis, we use a velocity boundary for WW candidates of 10\,km\,s$^{-1}$ and consider stars above 30\,km\,s$^{-1}$ to be RW candidates.

The ONC is thought to have produced high-mass RW stars in the past, most notably AE Aurigae (AE~Aur) and $\mu$ Columbae ($\mu$~Col), which were among the first identified runaways. \citet{RN325} showed that these two stars were moving in almost opposite direction from the ONC at space velocities of $\sim$100\,km\,s$^{-1}$ and suggested that they were ejected in the same event $\sim$2.6 Myr ago. \citet{RN50} used Monte-Carlo simulations to show that observations of these two stars are consistent with having originated in the Trapezium Cluster (at the centre of the ONC) and their runaway status being a consequence of a binary-binary dynamical ejection $\sim$2.5 Myr ago. 

The Becklin-Neugebauer (BN) object \citep{RN327} is another fast moving, high-mass star that has been postulated to have been recently ejected from the Orion region, however its exact origin is still debated \citep{RN328,RN330,RN329,RN326}. Unlike AE~Aur and $\mu$~Col, which are both visible in the optical spectrum the BN-object is an IR source and not visible in the wavelength range covered by \textit{Gaia}.

The ONC has also been suggested as the origin of three potential low-mass runaways \citep{RN227}. These candidates were identified based on their high proper motion and converted to tangential velocities (38--69\,km\,s$^{-1}$) based on an assumed distance of 470 pc. However, \citet{2005ApJ...633L..45O} used Hubble Space Telescope observations to show that these three stars do not actually move fast enough to be classified as RW stars. With these new velocities (5.5--7.9\,km\,s$^{-1}$) we would not even consider them as WW stars. Given the suggested upper limit escape speed \citep{RN322} of $\sim$5.8\,km\,s$^{-1}$, these stars might not be unbound from the ONC if this is in fact their birth region.

Recently, \citet{2019ApJ...884....6M} compiled a list of known young stellar objects from literature within a search radius of 2$\degree$ around the ONC and a parallax limit of 2 < $\varpi$ < 5 mas. They cross-matched these stars with \textit{Gaia} DR2 data and applied photometric and minimum proper motion cuts. These steps resulted in 26 potential RW candidates having been ejected from the ONC. After tracing these candidates back, they identify 9 stars with an apparent origin close to the Trapezium Cluster at the centre of the ONC. Our analysis covers this region as well and we will seek to confirm these candidates in our analysis.

\subsection{Data selection and filtering}

We adopt the centre of the ONC as described in \citet{RN264} with values as shown in Table~\ref{tab:ONC_para}. From this central location we select sources within a radius of 14$\degree$, which at the distance of the ONC centre translates to a radius of $\sim$100 pc. We apply the distances from the catalogue by \citet{RN305} to all sources within this radius and reduce the sample by selecting only sources with distances between 300 and 500 pc, adopting a central distance of 400~pc, instead of 403 pc from \citet{RN264}.
\begin{table}
	\centering
	\caption{ONC centre parameters used in the analysis; [1]~\citet{RN264}}
	\label{tab:ONC_para}
	\begin{tabular}{lr}
		\hline
		Right ascension [RA] (ICRS) $\alpha_{0}$ & 5h 35m 16s [1]\\ 
		Declination [Dec] (ICRS) $\delta_{0}$ & -05$\degree\,$23$^\prime\,$40$^{\prime\prime}$ [1]\\ 
		Proper motion RA ${\mu}_{{\alpha }^{\star},0}$ (mas\,yr$^{-1}$) & 1.51\,$\pm$\,0.11 [1]\\
		Proper motion Dec ${\mu}_{{\delta}_{0}}$ (mas\,yr$^{-1}$) & 0.50\,$\pm$\,0.12 [1]\\
		RV (\,km\,s$^{-1}$) &  21.8\,$\pm$\,6.6 [1]\\
		\hline
		Adopted distance (pc) & 400\\
		Adopted parallax $\varpi_{0}$ (mas) & 2.50\\
		\hline
%		\multicolumn{3}{l}{[1] , [2] some other survey, [3] }
	\end{tabular}
\end{table}

\begin{figure}
    \centering
    \includegraphics[width=1\linewidth]{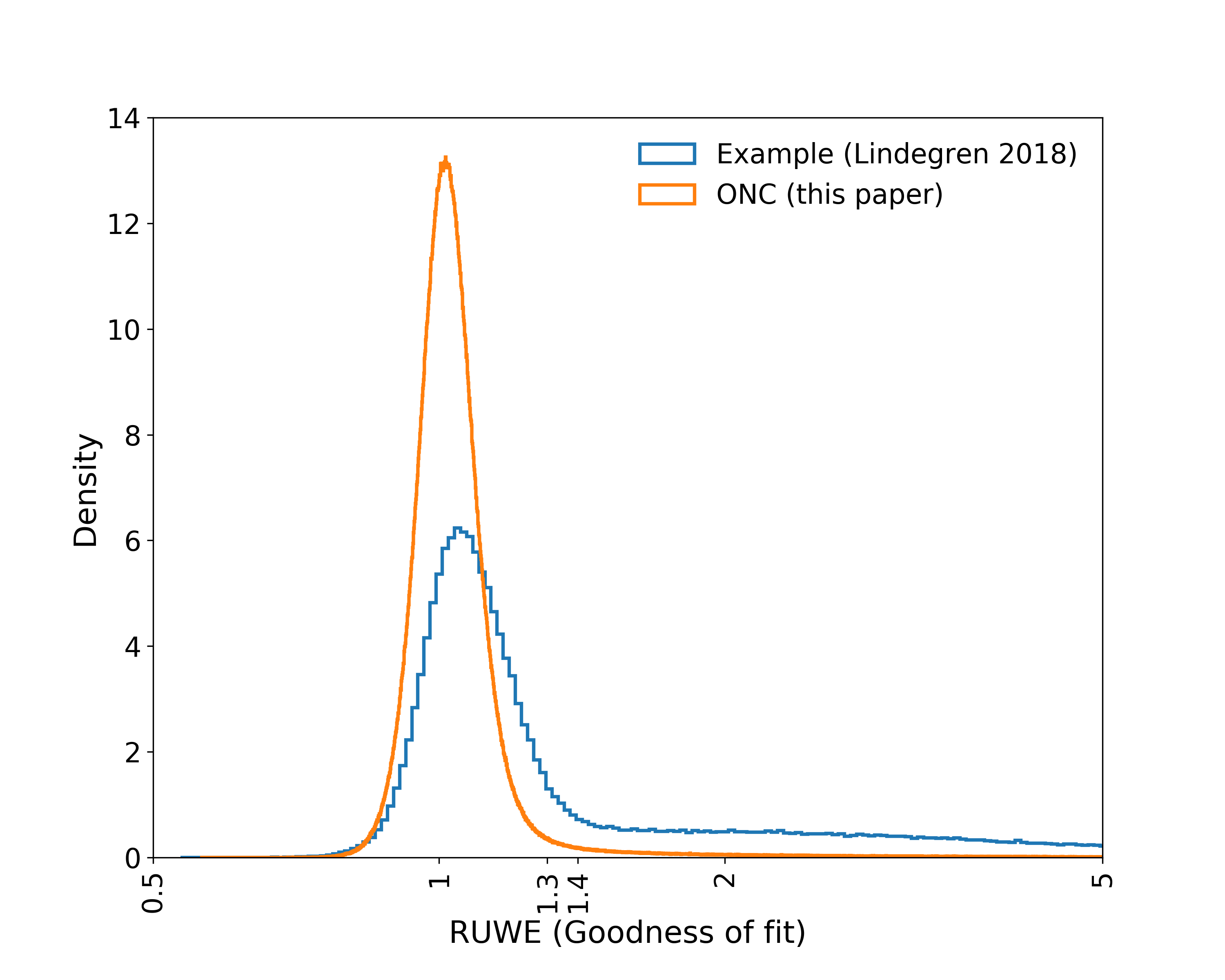}
    \caption{Distribution of our ONC data (orange line) according to RUWE value following the example approach in \citet{LL:LL-124}. The data from this example (blue line) results in a defined RUWE boundary of 1.4. The ``knee'' of our ONC distribution is located at RUWE = 1.3, which we use as our upper limit for the goodness of fit of observations with a single-star model.}
    \label{fig:RUWE_1.2}
\end{figure}

Cuts to increase astrometric and photometric quality are applied to this data. We use the re-normalised unit weight error RUWE \citep[][``\textit{Gaia} known issues'' website\footnote{https://www.cosmos.esa.int/web/gaia/dr2-known-issues}]{LL:LL-124}, which is an indicator for how well the single-star model fits to the observations. Higher RUWE values indicate problems with the astrometry or the presence of a non-single star. The technical note GAIA-C3-TN-LU-LL-124-01 \citep{LL:LL-124} provided an example for calculating an upper boundary for a good fit. This example derived a RUWE boundary of 1.4. Plotting our data in the same way as described in the technical note results in a RUWE distribution as shown in Fig.~\ref{fig:RUWE_1.2}, where we also include the example distribution from \citet{LL:LL-124}. The resulting graph shows that excluding stars with RUWE > 1.4 leaves stars located in the upper tail of the distribution. We consider RUWE = 1.3 a better (more conservative) upper boundary for our data set and exclude all data above this value.

The photometric excess noise (flux excess factor $E$) filter \citep{RN307,2018A&A...616A..10G}:
\begin{equation}\label{eq:Phot_excess}
    1.0 + 0.015(G_{\text{BP}}-G_{\text{RP}})^2 < E < 1.3 + 0.06(G_{\text{BP}}-G_{\text{RP}})^2
\end{equation}
\noindent provides us with high-quality photometric data as well as further cleaning up the astrometry \citep{2018A&A...616A..17A}. This filter removes sources with spurious photometry in dense, crowded areas (such as the centre of the ONC) \citep{RN307,2018A&A...616A..10G}. In our case, it filters a large number of stars located in the central region of the ONC, which does not greatly affect our analysis as we are mainly interested in stars that are no longer in this region.

\section{\textit{Gaia} DR2 Data analysis}

For our data analysis, we follow the approach as shown in \citet{RN264} and \citet{2019MNRAS.487.2977G}. The radial velocity (RV) of a young star-forming region (or older globular cluster) can have a perspective effect on the proper motion and create a perspective contraction (if RV > 0) or expansion (if RV < 0) \citep{2009A&A...497..209V,2018A&A...616A..12G}. This perspective effect in the proper motion ($\Delta{\mu }_{{\alpha }^{\star },i}$ and $\Delta{\mu }_{{\delta }_{i}}$) of each star $i$ in the region can be corrected for using a first-order approximation as described in equation 13 by \citet{2009A&A...497..209V}:
\begin{equation}
        \Delta{\mu }_{{\alpha }^{\star },i} \approx {\rm{\Delta }}{\alpha }_{i} \left({\mu }_{{\delta }_{0}} \sin{\delta_{0}} - \cfrac{V_{\text{rad,0}}\,\varpi_{0}}{\kappa} \cos{\delta_{0}} \right)\,,
\end{equation}
\begin{equation}
        \Delta{\mu }_{{\delta }_{i}} \approx - {\rm{\Delta }}{\alpha }_{i} {\mu }_{{\alpha }^{\star },0} \sin{\delta_{0}} - {\rm{\Delta }}{\delta }_{i} \cfrac{V_{\text{rad,0}}\,\varpi_{0}}{\kappa}\,.
\end{equation}

\noindent In these two equations $\alpha_{0}$, $\delta_{0}$, $\varpi_{0}$ represent the position and ${\mu }_{{\alpha }^{\star },0}$, ${\mu }_{{\delta }_{0}}$ and $V_{\text{rad,0}}$ the velocity parameters of the centre of the cluster. ${\rm{\Delta }}{\alpha }_{i}$ and ${\rm{\Delta }}{\delta }_{i}$ represent the differences between right ascension and declination of each star and the cluster centre. $\kappa$ is the conversion factor to convert RV from km\,s$^{-1}$ to mas\,yr$^{-1}$ at a distance of 1 kpc and has a value of 4.74. We apply this correction to the proper motion of all stars in our data set. We use the ONC central parameter values as shown in Table~\ref{tab:ONC_para}.

We then remove the Sun's peculiar motion relative to the Local Standard of Rest (LSR) \citep[using velocity values from][]{2010MNRAS.403.1829S} from the velocities of all stars in the data set as well as from the ONC central velocity parameters (see Table~\ref{tab:ONC_para}). We then apply a rest frame that is centred on the ONC to our data set by subtracting the central velocity parameters.

Finally, we transform the positions and proper motions into a Cartesian coordinate system using orthographic projection \citep[Eq. 2,][]{2018A&A...616A..12G}. We define the xy-plane as a projected representation of the positions on the sky and the z-direction as the radial direction. Radial velocities are only transformed into the rest frame by subtracting the central radial velocity without any further corrections or changes \citep[see][]{2019MNRAS.487.2977G}.

\subsection{Search procedure}\label{Search_proc}

Most of the stars ($\sim$93 per cent) in our data set do not have a measured RV in \textit{Gaia} DR2. As a consequence, we start with a search for 2D-candidates and only when confirmed as a 2D-candidate do we proceed in three dimensions for stars with RV. For 2D-candidates without RV-measurements in \textit{Gaia} DR2, we search the literature for RV and, where available, convert these into the ONC rest frame. 

We search for RW (velocity > 30\,km\,s$^{-1}$) and WW (velocity: 10--30\,km\,s$^{-1}$) 2D-candidates by tracing back their positions in the xy-plane for up to 4 Myr using the converted proper motions. Once this backwards path crosses the cluster boundary, a star becomes a candidate and the time at which it intersects the boundary becomes its minimum flight time, i.e. minimum time since ejection. 

The ONC is fairly well-constrained on the sky (position and size) allowing us to define an approximate cluster boundary by using the nominal radius of 2.5 pc \citep{RN321}. This corresponds to an angular size of $\sim$0.35\degree around the ONC centre position, which is located at the origin in the ONC rest-frame. However, the ONC is far less constrained in the radial direction.  Older estimates positioned it further away, e.g. \citet{2007A&A...474..515M} determined it to be at a distance of 414\,$\pm$7 pc. Recent estimates have since reduced the distance to the ONC, \citet{2018AJ....156...84K} derived a distance of 389\,$\pm$3 pc, whereas \citet{RN264} located it at 403\,$^{+7}_{-6}$ pc, both using \textit{Gaia} DR2. 

The size of the ONC in the line-of-sight direction is less constrained than it is on the plane of the sky. \citet{RN333} used \textit{Gaia} DR2 to investigate the 3D-shape of the Orion A region, which includes the ONC at its ``Head''. The authors suggest that the ONC region extends from its centre for about 15--20 pc in either direction. For our analysis, we consider our cluster boundary in the line-of-sight direction to extend 15 pc either direction from our adopted centre at 400 pc, which is located at the origin of our rest frame.

For 2D-candidates without RV, we use the radial distance of these candidates to the cluster boundary and each candidate's minimum flight time to calculate a required RV to reach this distance since ejection. If the resulting velocity is > |500|\,km\,s$^{-1}$ we exclude these 2D-candidates from the list of 2D-candidates. 

We have searched through several RV surveys, i.e. RAVE DR5 \citep{2017AJ....153...75K}, GALAH DR1 \citep{2017MNRAS.465.3203M} and also the Simbad/VizieR databases \citep{2000A&AS..143....9W, 2000A&AS..143...23O} to complete our data set with secondary RV measurements. We find additional RVs in \citet{2006AstL...32..759G}, \citet{2015ApJ...807...27C}, and \citet{2018AJ....156...84K} for several 2D RW and WW candidates. We also find secondary, more precise RVs for several sources, where the \textit{Gaia} DR2 RVs have large errors, and use these secondary RVs instead.

Before tracing back 2D-candidates with RVs in three dimensions, we exclude those 2D-candidates where the RV points towards the ONC as these stars cannot have originated from the ONC. We then trace back the remaining candidates in the same way as described for the xy-plane in the xz-plane and yz-plane. Instead of a circular cluster boundary, the boundary in the latter two planes turns into an ellipse with semi-minor axis of 2.5 pc (in the x and y axes) and semi-major axis of 15 pc (in the z axis).

\begin{figure*}
    \centering
    	\includegraphics[width=1\linewidth]{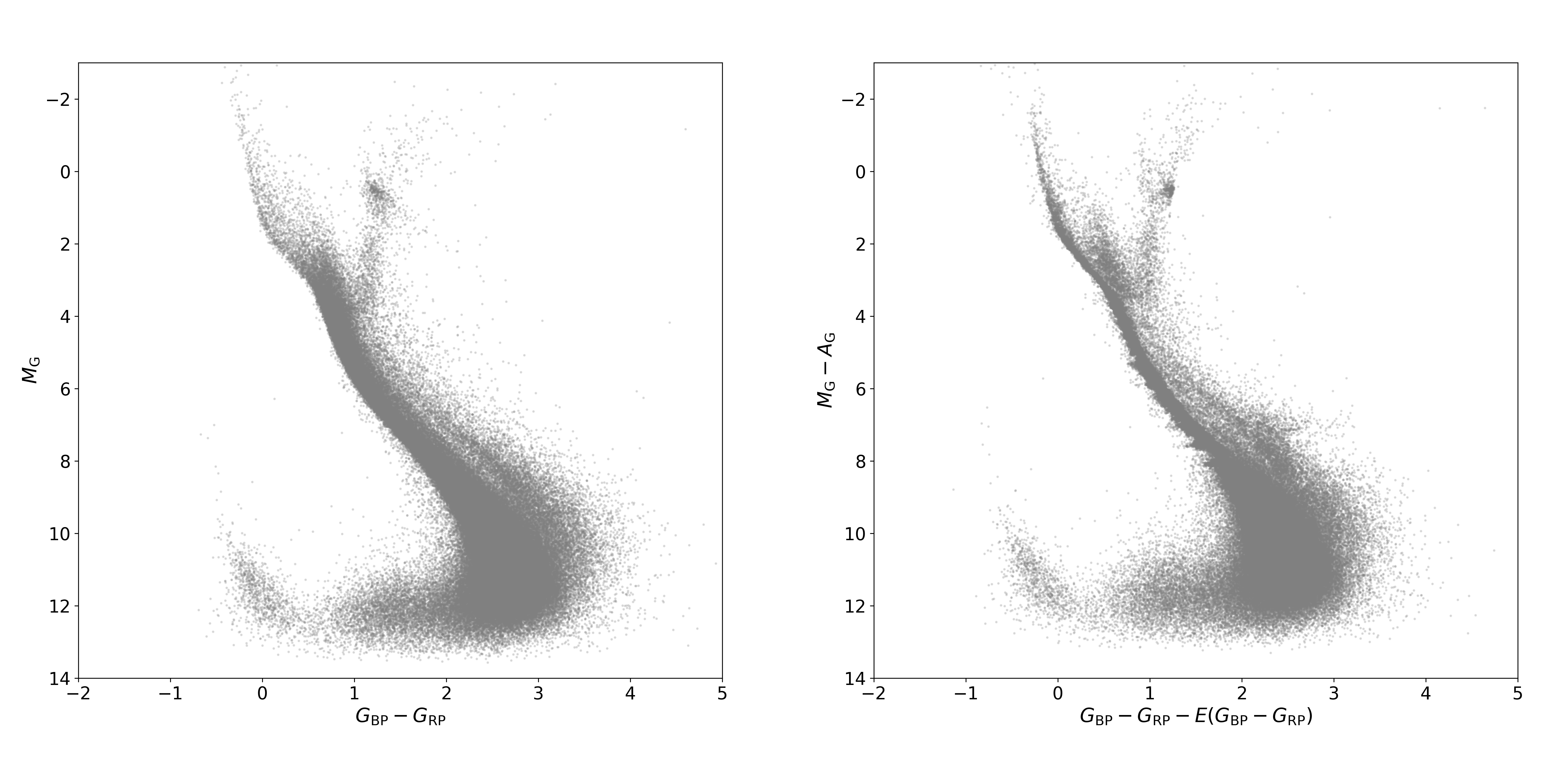}
        \caption{Left: Data before extinction and reddening correction, after applying astrometric and photometric filter. This CMD clearly shows at least two differently aged populations of stars with an obvious main-sequence and at least one younger population of stars above it. Right: ONC after correction. The correction sharpens the main sequence with a younger population of stars visible above it. Note: the horizontal stripes along the main sequence (in particular for $M_{\rm{G}}-A_{\rm{G}}$ between 6 and 9 mag) are artefacts and present only for stars with \textit{Gaia} DR2 extinction and reddening \citep{RN303}.}
        \label{fig:Extinction_reddening}
\end{figure*}

\subsection{Constructing the CMD}\label{CMD_constr}

Not all of the stars that we can trace back successfully to the ONC have originated there. As the ONC is less than 4 Myr old, RW/WW-candidates that are older than this might have passed through the ONC during its existence, but were not born there. To further narrow down the list of candidates based on age, we construct CMDs with those stars that can be successfully traced back in 2D and 3D.

As the data set covers distances between 300--500 pc, we estimate the absolute $G$-band magnitude of each star $M_{\rm{G}}$ using its apparent magnitude $G$ and distance $r$ \citep{RN305} in equation 2 in \citet{RN303}:
\begin{equation}\label{Distance_Modulus}
    M_{\rm{G}} = G +5 - 5 \log_{10} r - A_{\rm{G}}\,,
\end{equation}

\noindent which also includes a correction for the extinction in the $G$-band, $A_{\rm{G}}$. In the CMDs, we plot this extinction-corrected $M_{\rm{G}}$ against $G_{\rm{BP}}-G_{\rm{RP}}$, which we correct for reddening, $E(G_{\rm{BP}}-G_{\rm{RP}})$ as shown in Fig.~\ref{fig:Extinction_reddening}.

\subsubsection{Extinction and reddening correction}

Extinction, $A_{\rm{G}}$ and reddening, $E(G_{\rm{BP}}-G_{\rm{RP}})$ values are available only for a subset of all sources in the \textit{Gaia} DR2 catalogue. To estimate the missing correction values, we follow a similar approach to \citet{2018A&A...620A.172Z}. These authors assigned missing values based on the 3D-position of a star using averages of \textit{Gaia} DR2 extinction and reddening values from surrounding stars. Using the transformed Cartesian coordinates for position and distance, we draw a 10 pc sphere around each star with missing values, calculate the average $A_{\rm{G}}$ and $E(G_{\rm{BP}}-G_{\rm{RP}})$ values in this sphere from sources with \textit{Gaia} DR2 values and assign these averages to the star in the centre. The effect of this correction method of on our complete data set is shown in Fig.~\ref{fig:Extinction_reddening}.

In the right panel of Fig.~\ref{fig:Extinction_reddening}, we see the main sequence (MS) emerging more clearly compared to the left panel and a secondary brighter track above it. This secondary track will include stars younger than the MS, however it can also include older binaries. Our choice of a low RUWE-value should reduce the likelihood of older binaries contaminating this region. We cannot fully remove the risk as the RUWE filter will only remove binaries, where the astrometric quality is compromised by the binary status, e.g. systems with larger orbital periods \citep{2020arXiv200305456P,2020arXiv200305467B}. 

In this right panel we also see horizontal stripes along the MS, most obvious at an absolute magnitude of 6 to 9 mag. These are a result of the way the PARSEC evolutionary tracks were sampled in \textit{Gaia} DR2 and are artefacts \citep{RN303}. This affects only stars with \textit{Gaia} DR2-provided $A_{\rm{G}}$ and $E(G_{\rm{BP}}-G_{\rm{RP}})$ values and not those calculated by our method of averaging over neighbouring stars.

\subsubsection{Age estimate using PARSEC isochrones}

We use an upper age limit of 4 Myr and consider only stars that are younger than this age to have possibly originated in the ONC. To get age estimates of our candidate stars, we use PARSEC isochrones \citep[version 1.2S,][]{RN225} to separate the stars into two age brackets (younger stars plotted above the isochrone, older stars below). We download data\footnote{http://stev.oapd.inaf.it/cgi-bin/cmd} to produce an isochrone using a linear age of 4 Myr with a mean metallicity of $Z$ = 0.011268 \citep{RN331,RN232} and interstellar extinction $A_{\rm{V}}$ = 0. We use the \citet{2018A&A...616A...4E} passbands. 

\subsection{Error calculation}

\subsubsection{Astrometric errors}\label{Astrometric error}

To describe the velocity errors for our RW and WW candidates, we follow a similar approach to \citet{RN264}. We use a covariance matrix based on the \textit{Gaia} DR2 covariance matrix for the astrometric solution using equation B.3 from \citet{RN307} and convert into errors in our Cartesian coordinate system. We then multiply these internal \textit{Gaia} DR2 uncertainties with a correction factor of 1.1 \citep{RN307}, however we do not consider any corrections for systematic errors in the proper motions. Finally, we convert the errors from proper motions into velocities using distances $r$ and $\kappa$=4.74 (conversion factor from mas\,yr$^{-1}$ to km\,s$^{-1}$).

\subsubsection{Photometric errors}\label{Photometric error}
We calculate photometric errors in G-magnitude, $G_{\rm{BP}}$ and $G_{\rm{RP}}$. No errors are provided for these quantities in \textit{Gaia} DR2 as the error distribution is only symmetric in flux space \citep{2018gdr2.reptE..14H}. This converts to an asymmetric error distribution in magnitude space which cannot be represented by a single error value. The G-magnitude in \textit{Gaia} DR2 is calculated following equation 5.20 in the Gaia DR documentation \citep{2018gdr2.reptE...5B} adding a zero point, $G_{\rm{0}}$ to the instrumental $G$-magnitude value:
\begin{equation}
    G = G_{\rm{instr}} + G_{\rm{0}}
\end{equation}
Using equation 5.26 in the Gaia DR documentation \citep{2018gdr2.reptE...5B} allows us to calculate the error in $G$:
\begin{equation}
    \sigma_G = \sqrt{\left(1.0857 \cfrac{\sigma_{\,\bar{I}}}{\bar{I}} \right)^2 + \left(\sigma_{G_{\rm{0}}} \right)^2}
\end{equation}
\noindent In this equation $\sigma_{\,\bar{I}}$ represents the error from internal calibration, labeled in the data as \texttt{phot\_g\_mean\_flux\_error} and $\bar{I}$ represents the weighted mean flux, labeled in the data as \texttt{phot\_g\_mean\_flux} in \textit{Gaia} DR2. $\sigma_{G_{\rm{0}}}$ represents the passband error in the zero point.

We calculate the errors for $G$-magnitude, $G_{\rm{BP}}$ and $G_{\rm{RP}}$ using the passband errors in the zero points in the VEGAMAG system \citep{2018A&A...616A...4E}. We then transform the apparent $G$-magnitude errors into absolute $M_{\rm{G}}$-errors using distances from \citet{RN305}, also considering errors in these distances and calculate the errors for $G_{\rm{BP}} - G_{\rm{RP}}$ and include errors in extinction and reddening. 

For stars with \textit{Gaia} DR2 data for extinction $A_{\rm{G}}$ and reddening $E(G_{\rm{BP}}-G_{\rm{RP}})$, we are also provided with upper and lower percentile values, which we use to calculate upper and lower errors. For stars with averaged correction values we take the standard deviation of the values that we averaged over to calculate extinction and reddening errors. The final photometric errors are dominated by errors in extinction, $A_{\rm{G}}$ and reddening, $E(G_{\rm{BP}}-G_{\rm{RP}})$ with smaller contributions from errors in distance, apparent magnitude and colour.

\section{\textit{N}-body simulations of the ONC}
To predict the number and velocities of RW and WW stars in the observational data, we have run a set of 20 $N$-body simulations with initial conditions similar to those the ONC is thought to have evolved from. \citet{RN309} used $N$-body simulations to show that the number and velocity distribution of ejected stars can be used to constrain the initial spatial and kinematic substructure and we use this approach to provide predictions for our search.

\subsection{Simulation set-up}
The ONC is thought to have evolved from an initial state of being spatially and kinematically substructured \citep{RN4, RN38}. Spatial substructure can be created in $N$-body simulations by using fractal distributions, as shown in \citet{RN14}. The degree of substructure is defined using only a single parameter, the fractal dimension $D$. In our $N$-body simulations, we use a fractal dimension $D$ = 2.0, which produces a moderate amount of spatial substructure. The use of fractals also allows us to set up the initial kinematic substructure. Velocities of stars that are close to each other are correlated, but more distant stars can have very different velocities \citep{RN14}. The velocities in our simulations are scaled so the regions are initially subvirial with a virial ratio $\alpha_{\text{vir}}$ = 0.3, where $\alpha_{\text{vir}}$ = $ T/|\varOmega|$, with $T$ as the total kinetic energy and $|\varOmega|$ the total potential energy of the stars. For a detailed description of the construction of the fractals in the simulations, we refer to \citet{RN14} and also \citet{RN5, RN1}.

We use a larger number of systems than \citet{RN309}, i.e. 2000 systems compared to 1000 systems per simulation to reflect the higher number of stars in the ONC. The masses for the systems are sampled randomly from a \citet{RN203} IMF with stellar masses between 0.1 M$_{\sun}$ and 50 M$_{\sun}$. This upper mass limit of our sample is consistent with the mass estimate of the most massive star in the ONC, which is $\theta^1$ Ori C, a visual binary system with a total mass of $\sim$50 M$_{\sun}$ and a $\sim$30--35 M$_{\sun}$ primary \citep[e.g.][]{1998ApJ...492..540H, 2007A&A...466..649K,2008hsf1.book..483M,2010A&A...514A..34L}.

The Maschberger IMF is a combination of a Chabrier (\citeyear{RN200}) lognormal IMF approximation for low-mass stars combined with the power-law slope of Salpeter (\citeyear{RN204}) for stars above 1 M$_{\sun}$:
\begin{equation}
    p(m) \propto \cfrac{\left(\cfrac{m}{\mu}\right)^{-\alpha}}{\left(1+\left(\cfrac{m}{\mu}\right)^{1-\alpha}\right)^\beta}
    \label{eq:MaschbergerIMF}
\end{equation}
\noindent This IMF \citep{RN203} is described by a probability density function, where $\alpha$ = 2.3 (power-law exponent for higher mass stars), $\beta$ = 1.4 (describing the IMF slope for lower-mass stars) and $\mu$ = 0.2 (average stellar mass). 

We incorporate primordial binaries in our simulations with the binary fraction $f_{\text{bin}}$ depending on the mass of the primary star $m_{\text{p}}$. This fraction is defined as:
\begin{equation}
   f_{\text{bin}} = \frac{B}{S+B}
\end{equation}
\noindent S and B represent the number of single or binary systems in the simulations, respectively. We do not include any primordial higher-order multiple systems (triples or quadruples). We apply $f_{\text{bin}}$ as shown in Table~\ref{tab:Bin_frac} depending on the primary star's mass and with binary separations as shown in Table~\ref{tab:Bin_sep}. These binary fractions and separations are similar to the Galactic field, as the ONC is thought to be consistent with it \citep{2006A&A...458..461K,2007AJ....134.2272R,RN8, RN257}.
\begin{table}
	\centering
	\caption{Binary fractions in the $N$-body simulations. Column 1 shows the mass range based on the mass of the primary star; Column 2 shows the binary fraction $f_{\text{bin}}$.}
	\label{tab:Bin_frac}
	\begin{tabular}{lcl} 
		\hline
		$m_{\text{p}}$ [M$_{\sun}$] & $f_{\text{bin}}$ & Source \\
		\hline
		0.10 $\leq m_{\text{p}}$  < 0.45 & 0.34 & \citet{2012ApJ...758L...2J}\\
		0.45 $\leq m_{\text{p}}$  < 0.84 & 0.45 & \citet{1992ASPC...32...73M}\\
		0.84 $\leq m_{\text{p}}$  < 1.20 & 0.46 & \citet{2012ApJ...745...24R}\\
		1.20 $\leq m_{\text{p}} \leq$ 3.00 & 0.48 & \citet{2012MNRAS.422.2765D,2014MNRAS.437.1216D}\\
		$m_{\text{p}}$  > 3.00 & 1.00 & \parbox[t]{4cm}{\citet{1998AJ....115..821M}\\\citet{2007AA...474...77K}} \\		
		\hline
	\end{tabular}
\end{table}

\begin{table}
	\centering
	\caption{Mean binary separations in the $N$-body simulations. Column 1 shows the mass range based on the mass of the primary star; Column 2 shows the mean binary separation $\bar{a}$; Column 3 represents the variance $\sigma_{\textrm{log}\,\bar{a}}$ of the log-normal fit to the binary separation distributions.}
	\label{tab:Bin_sep}
	\begin{tabular}{lccl} 
		\hline
		$m_{\text{p}}$ [M$_{\sun}$] & $\bar{a}$ [au] & $\sigma_{\textrm{log}\,\bar{a}}$ & Source \\
		\hline
		0.10 $\leq m_{\text{p}}$ < 0.45 & 16 & 0.80 & \citet{2012ApJ...758L...2J} \\ 
		0.45 $\leq m_{\text{p}}$ < 1.20 & 50 & 1.68 & \citet{2012ApJ...745...24R}\\
		1.20 $\leq m_{\text{p}} \leq$ 3.00 & 389 & 0.79 & \citet{2014MNRAS.437.1216D}\\
		$m_{\text{p}}$ > 3.00 & \parbox[t]{1.01cm}{\centering \"Opik~law\\(0-50)}& -- & \parbox[t]{2.8cm}{\citet{1924PTarO..25f...1O},\\\citet{2013AA...550A.107S}} \\
		\hline
	\end{tabular}
\end{table}

\noindent For stars in primordial binaries, the secondary star is assigned a mass $m_{\text{s}}$ based on a flat mass ratio distribution, which is observed in the field and many star-forming regions \citep[e.g.][]{2011ApJ...738...60R, 2013A&A...553A.124R}. 
The binary mass ratio $q$ is: 
\begin{equation}
    q = \cfrac{m_{\text{s}}}{ m_{\text{p}}}\,.
\end{equation}
In our simulations we allow this secondary to have a mass of $m_{\text{p}}$ > $m_{\text{s}}\geq$ 0.01 M$_{\sun}$, so the secondary star can also be a brown dwarf (BD). These binary fractions result in an average total number of stars in our simulations of $\sim$2800 stars, average cluster masses of $\sim$2100 M$_{\sun}$ and an escape velocity of $\sim$6\,km\,s$^{-1}$, which is consistent with the estimated total cluster mass in \citet{1998ApJ...492..540H} and the escape velocity estimate in \citet{RN322}.

\begin{figure*}
    \centering
    \begin{minipage}[t]{1.0\columnwidth}
        \centering
        \vspace{0pt}
    	\includegraphics[width=1.0\linewidth]{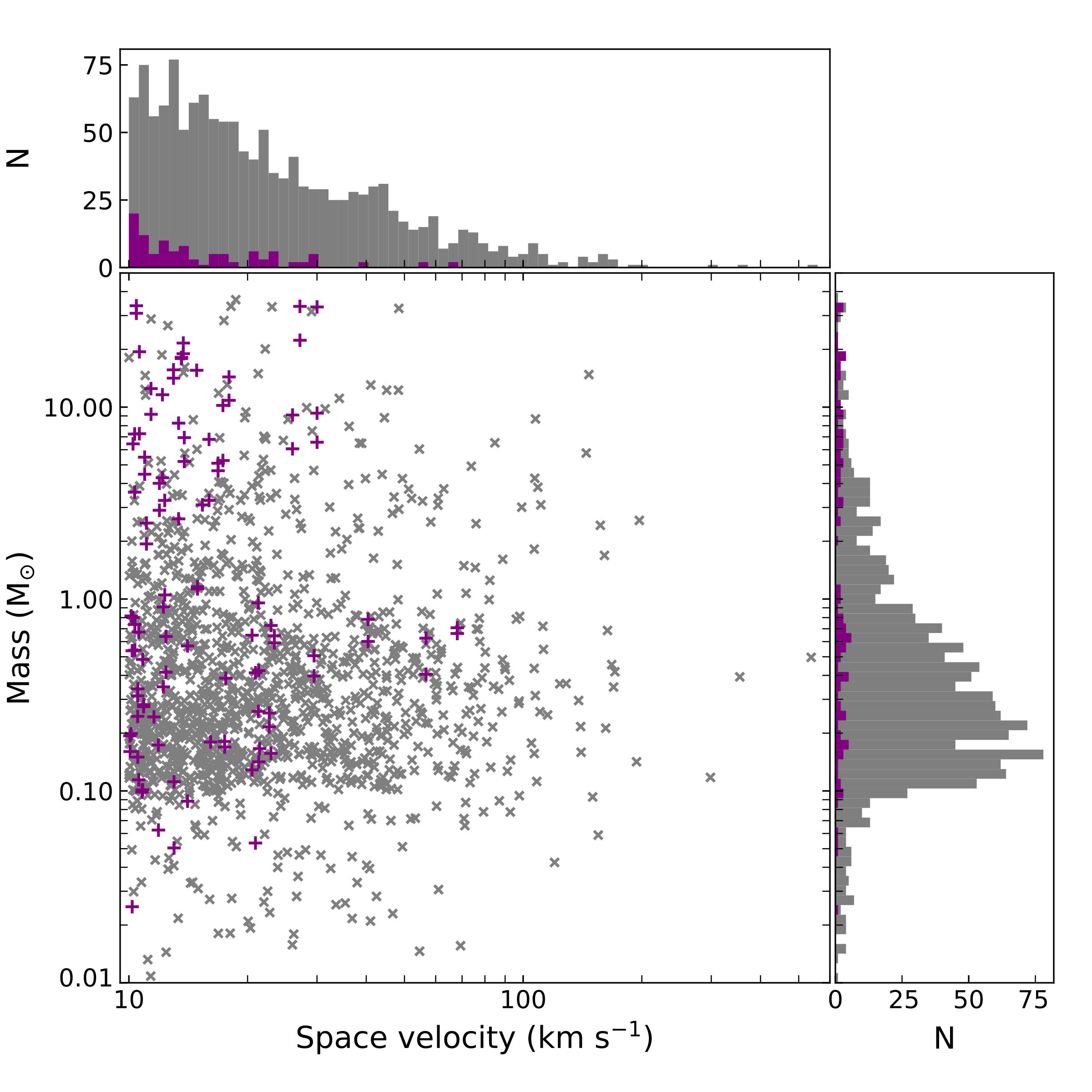}
    \end{minipage}
    \hfill{}
    \begin{minipage}[t]{1.0\columnwidth}
        \centering
        \vspace{0pt}
        \includegraphics[width=1.0\linewidth]{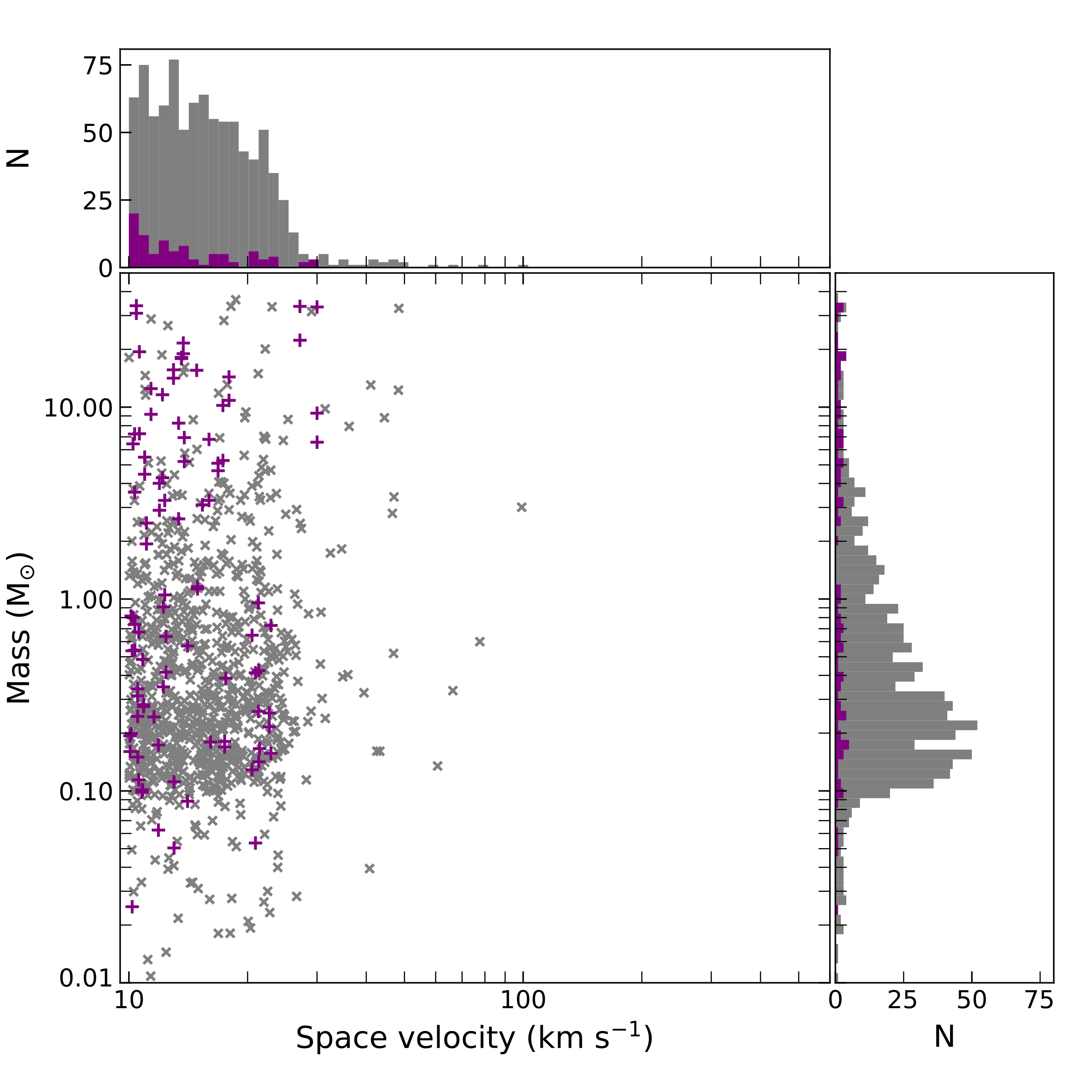}
    \end{minipage}
    \caption{Masses and velocities of all RW/WW stars from the 20 simulations after 4 Myr. Stars with masses \textit{m} < 0.1 M$_{\sun}$ are current or previous brown-dwarf binary companion stars. Ejected binaries are marked with a purple ``+'', whereas single stars are marked with a grey ``x''. Left panel: RW/WW stars at all distances. Right panel: RW/WW stars still located within the 100 pc search boundary.}
    \label{fig:mass_vel_plot}
\end{figure*}

The initial binary separation, i.e. semi-major axis, is based on a log-normal distribution with mean values for the separation $\bar{a}$ in astronomical units (au) and the variance shown in Table~\ref{tab:Bin_sep} for different primary masses. These values also follow recent observations of binaries in the field \citep[e.g.][]{2013ARA&A..51..269D,2014MNRAS.442.3722P}. The initial eccentricities of binaries with orbital periods <~0.1~au have circular orbits, which is in line with observations. Binaries with larger orbital periods have eccentricities drawn randomly from a flat distribution \citep[e.g.][]{2013ARA&A..51..269D,2014MNRAS.442.3722P,2019MNRAS.485L..48W}. For further information on the set-up of the binary systems, see \citet{2014MNRAS.442.3722P}.

We use the $N$-body integrator {\texttt kira} and the stellar and binary evolution package {\texttt SeBa} from the {\texttt Starlab} environment \citep{RN236,RN193}. We evolve our star-forming regions over a defined time period of 4 Myr and take snapshots every 0.01 Myr. The initial radius of our star-forming region is 1.5 pc and we have not applied any external tidal field.

The stellar systems in our simulations undergo stellar and binary evolution. We do not see any supernovae during our 4 Myr simulations. The furthest evolution step for the highest-mass stars is that of a helium star. We do however see several binary mergers as a result of binary evolution. 

\subsection{Predictions from the simulations}

The left panel of Fig.~\ref{fig:mass_vel_plot} shows the masses and space velocities of all ejected stars after 4 Myr from 20 $N$-body simulations that reach at least WW velocities (> 10\,km\,s$^{-1}$) at time of ejection. The distribution highlights that the highest velocities are achieved by lower-mass stars and that we should find runaway and walkaway stars across the mass spectrum around the ONC.

Most of the RW and WW stars are ejected with a space velocity of < 200\,km\,s$^{-1}$, however we have 3 sub-solar mass RW stars travelling with velocities between $\sim$300--540\,km\,s$^{-1}$. These RW-velocities are far above our average; however they are not improbable for the dynamical ejection scenario \citep[e.g.][]{1990AJ.....99..608L,RN54,2012ApJ...751..133P}. 

The fastest RW ($\sim$540\,km\,s$^{-1}$) from our simulations is the result of multiple dynamical interactions starting with a binary-binary interaction between two of the primordial binaries. The fastest RW is the primary (P1, 0.5 M$_{\sun}$) in an almost equal-mass binary ($q$~=~0.85). After 0.01 Myr, this binary interacts with another binary with a low mass ratio $q$ = 0.03 and a primary (P2) of 24 M$_{\sun}$. This interaction leads to the ejection of the secondary (S1, 0.4 M$_{\sun}$) from the equal-mass binary, turning S1 into a WW star. The now single P1 replaces the secondary (S2, 0.8 M$_{\sun}$) in the unequal mass binary. The system continues as a triple system with S2 turning into a tertiary companion on a wider orbit around the close binary P2-P1.

This triple system moves towards the region's centre as the region collapses due to its initial subvirial ratio. The tertiary S2 gets ejected at 0.5 Myr as a WW after further dynamical interactions. The remaining binary forms short-lived dynamical multiples with different stars until it gets fully disrupted at 3 Myr by an encounter with a 10 M$_{\sun}$ star. Our fastest star P1 gets ejected with a velocity close to its previous orbital velocity, whereas the high-mass primary P2 forms a new dynamical binary with the disrupting star and becomes an unbound binary just above the escape velocity. 

We also find ejected RW/WW binaries across the sampled mass range and these stars are highlighted with a purple ``+'' in Fig.~\ref{fig:mass_vel_plot}. The number of RW/WW binaries is low compared to the binary fractions of systems in the cluster. This is consistent with \citet{1990AJ.....99..608L} who showed that $\simeq$10 per cent of their ejected stars with velocities >30\,km\,s$^{-1}$ are binaries, compared to an initial binary fraction of 50 per cent. Also, \citet{2012ApJ...751..133P} suggest that in general the binary frequency of runaway stars is lower than that of the stars still within the cluster.

The maximum velocity of the ejected binaries in our simulation ($\sim$70\,km\,s$^{-1}$) is lower than for single stars, which is also consistent with the results of \citet{1990AJ.....99..608L} and \citet{2012ApJ...751..133P}. \citet{RN21} suggested that an ejected binary, resulting from a binary-single star encounter, receives less kinetic energy and travels at lower velocities than ejected single stars.

The stars in our simulations are sampled from the IMF down to 0.1 M$_{\sun}$, however we allow BD binary companions below this mass, so we also find a small number of BD RW and WW stars. All of our RW-BDs are single stars, however we have an occasional WW-BD ejected in a binary.

\begin{table*}
	\centering
	\caption{Ejected RW and WW stars from $N$-body simulations at all distances and within the search radius of 100 pc at different times during the simulations. We show the averages from all 20 simulations and maxima in a single simulation. [] indicates the number of RW/WW stars within 100 pc. We count ejected binary systems as one star when calculating averages and maxima. The uncertainties in our averages are the standard deviations.}
	\label{tab:RW_WW_pred}
	\begin{tabular}{lcc} 
		\hline
		Mass $m$ (M$_{\sun}$) & RW average / maximum & WW average / maximum \\
		\hline
		0.01 $\leq$ \textit{m} < 0.10  \\
		- after 1 Myr &  1.6$\,\pm$1.1 [1.5$\,\pm$1.1] / 4 [4] & 3.5$\,\pm$2.0 [3.5$\,\pm$2.0] / 7 [7] \\ 
		- after 2 Myr &  1.7$\,\pm$1.1 [1.0$\,\pm$1.1] / 4 [3] & 3.7$\,\pm$2.1 [3.7$\,\pm$2.1] / 7 [7] \\ 
		- after 3 Myr &  1.8$\,\pm$1.1 [0.3$\,\pm$0.4] / 4 [1] & 3.8$\,\pm$1.9 [3.8$\,\pm$1.9] / 6 [6]  \\ 
		- after 4 Myr &  1.8$\,\pm$1.1 [0.1$\,\pm$0.2] / 4 [1] & 3.8$\,\pm$1.9 [3.4$\,\pm$1.6] / 7 [6] \\ 
		0.10 $\leq$ \textit{m} < 8.00 \\
		- after 1 Myr &  15.5$\,\pm$4.3 [14.2$\,\pm$4.2] / 24 [22] & 38.3$\,\pm$5.1 [38.3$\,\pm$5.1] / 46 [46]\\ 
		- after 2 Myr &  16.7$\,\pm$4.1 [10.8$\,\pm$3.0] / 25 [15] & 41.7$\,\pm$5.0 [41.7$\,\pm$5.0] / 55 [55] \\ 
		- after 3 Myr &  17.2$\,\pm$4.5 [4.1$\,\pm$2.5]] / 27 [10] & 44.9$\,\pm$5.6 [44.9$\,\pm$5.6] / 57 [57] \\ 
		- after 4 Myr &  17.6$\,\pm$4.4 [1.0$\,\pm$1.0] / 26 [3] & 45.6$\,\pm$5.0 [41.6$\,\pm$5.5] / 57 [55] \\ 
		\textit{m} $\geq$ 8.00  \\
		- after 1 Myr &  0.2$\,\pm$0.4 [0.1$\,\pm$0.3] / 1 [1*] & 0.6$\,\pm$0.9 [0.6$\,\pm$0.9] / 3 [3] \\ 
		- after 2 Myr &  0.2$\,\pm$0.4 [0.1$\,\pm$0.3] / 1 [1*] & 0.8$\,\pm$1.1 [0.8$\,\pm$1.1] / 4 [4] \\ 
		- after 3 Myr &  0.3$\,\pm$0.5 [0.2$\,\pm$0.4] / 1 [1*] & 1.5$\,\pm$1.7 [1.5$\,\pm$1.7] / 6 [6] \\ 
		- after 4 Myr &  0.5$\,\pm$0.6 [0.3$\,\pm$0.5] / 2 [1*] & 1.7$\,\pm$1.5 [1.6$\,\pm$1.4] / 5 [5] \\ 
        \hline
		\multicolumn{3}{l}{\parbox[t]{9cm}{*depending on the simulation, we find 1 RW either within or outside of the 100 pc boundary}}
	\end{tabular}
\end{table*}

The right panel in Fig.~\ref{fig:mass_vel_plot} shows that the velocity distribution of RW/WW stars still within 100 pc of the cluster centre at the end of our simulations is different. Most of the WW stars are still within this radius, however most RW stars have passed through this region already. The maximum RW velocity drops to $\sim$100\,km\,s$^{-1}$ when only considering stars within the 100 pc region. 

Table~\ref{tab:RW_WW_pred} provides the average numbers of RW and WW stars across the mass ranges at four different times during the simulations and also gives the maximum we found in a single simulation. Most of our ejected stars are low/intermediate-mass stars (0.1 M$_{\sun}$ < $m$ < 8 M$_{\sun}$) with an average of 17.6$\,\pm$4.4 RW and 45.6$\,\pm$5.0 WW stars ejected at all distances from 20 simulation after 4 Myr. The maximum number of RWs in this mass range from a single simulation is 26, whereas we find a maximum of 57 WW stars.

When we limit ourselves to ejected low/intermediate-mass stars still within the 100 pc search region, we only find an average of 1.0$\,\pm$1.0 RW star and a maximum of 3 RWs after 4 Myr. At earlier times (younger ages) in our simulations, we see the number of RWs still within the search region increasing to an average of 14.2$\,\pm$4.2 and maximum of 22 at 1 Myr.

\begin{figure*}
    \centering
	\includegraphics[width=0.95\linewidth]{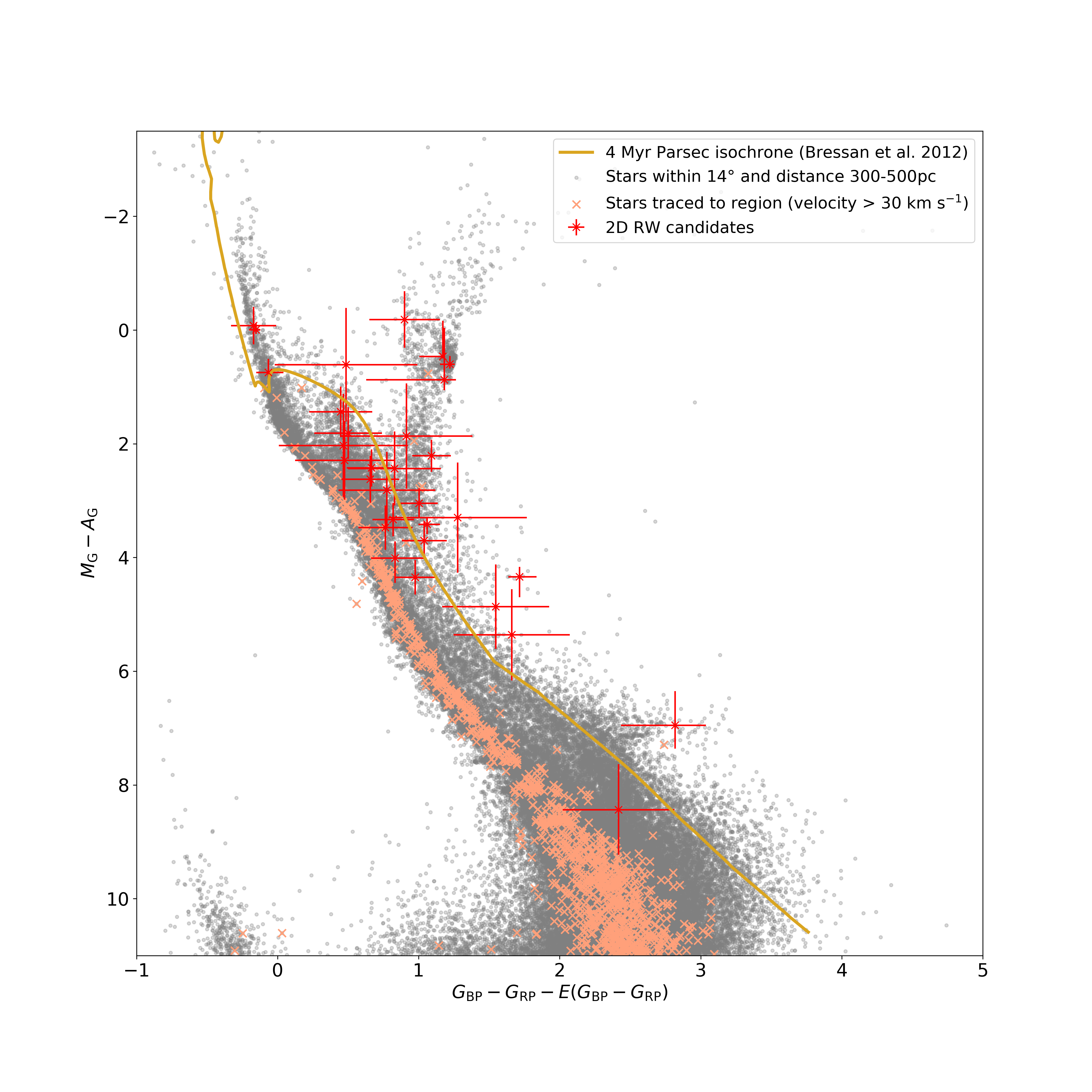}
    \caption{Colour magnitude diagram showing all 2D RW candidates (> 30\,km\,s$^{-1}$) that can be traced back to the ONC (red ``x''). We magnitude-limit the diagram to -3.5\,mag < $M_{\rm{G}}-A_{\rm{G}}$ < 11\,mag, which corresponds to a $G$-magnitude $\approx$ 19\,mag at the fainter end. Around this apparent magnitude value, the typical uncertainties in the 5-parameter astrometry increase quickly. The CMD includes a large number of stars that we have traced back to the ONC but that sit along the MS underneath the 4 Myr isochrone (orange ``x'') even when considering their errors, which are not plotted here. These stars are therefore too old to have been born in the ONC. We see a number of 2D-candidates at different absolute magnitudes that correspond to different masses. Many of our identified candidates have large errors in magnitude and colour, which are predominantly driven by the errors in the extinction and reddening. Some of these candidates sit below the 4 Myr isochrone, but might be younger than their position suggests due to the large errors.}
    \label{fig:2D_RW_candidates_after_correction}
\end{figure*}

At lower WW velocities we find that all ejected stars remain within the search region up to 3 Myr. Even after 4 Myr, we still find most WW stars, i.e. 41.6$\,\pm$5.5 stars (average) and 55 stars (maximum), within 100 pc. These findings suggest that we should find very few ejected RW stars, but most of the WW stars within our chosen 100 pc search radius around the ONC.

The number of RW/WW stars from the other two mass ranges (BDs and massive stars) are much lower. We only find a maximum of 2 high-mass RW stars (0.5$\,\pm$0.6 stars average) and 5 high-mass WW stars (1.7$\,\pm$1.5 stars average) at all distances. All of the WW stars at these masses are still located within the 100 pc search region, whereas only one of the two RW stars is. We also provide the number of ejected BDs in Table~\ref{tab:RW_WW_pred}, however do not include them in the following analysis as we are unlikely to be able to observe ejected BDs at the distance of the ONC. 

Most of the RW and WW stars are ejected as single stars. In different simulations, we find a maximum of one RW-binary and two WW-binaries composed of higher-mass stars. We also have a maximum of three RW-binaries and three WW-binaries composed of low/intermediate-mass stars (0.1 M$_{\sun}$ < $m$ < 8 M$_{\sun}$).

\begin{figure*}
        \centering
        \includegraphics[width=0.95\linewidth]{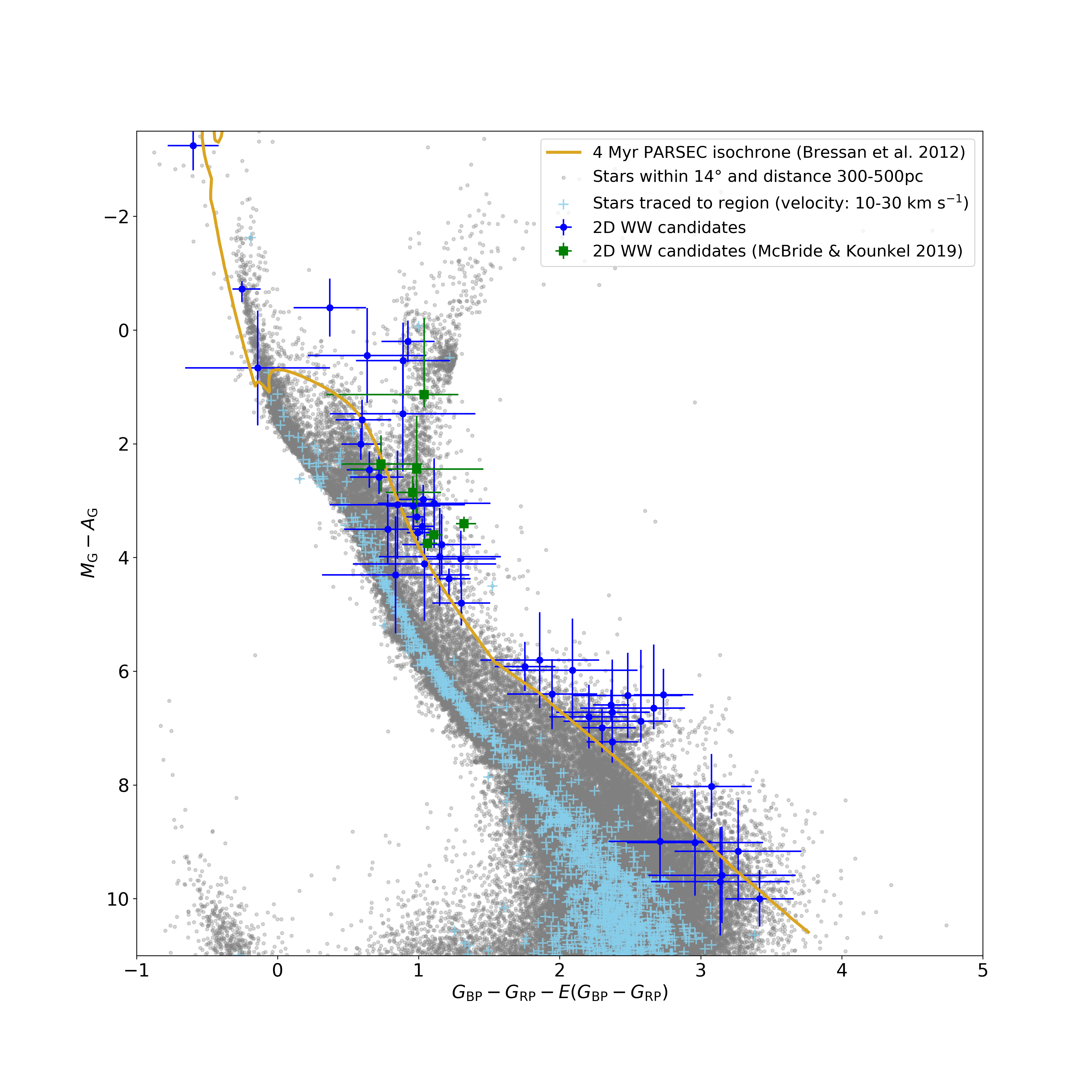}
        \caption{Colour magnitude diagram showing all 2D WW candidates (velocity: 10--30\,km\,s$^{-1}$) that can be traced back to the ONC (blue dot). We magnitude-limit the diagram to -3.5\,mag < $M_{\rm{G}}-A_{\rm{G}}$ < 11\,mag, which corresponds to a $G$-magnitude $\approx$ 19\,mag at the fainter end. Around this apparent magnitude value, the typical uncertainties in the 5-parameter astrometry increase quickly. The CMD includes a large number of stars that we have traced back to the ONC but that sit along the MS underneath the 4 Myr isochrone (light-blue ``+''). These stars are therefore too old to have been born in the ONC, even when considering their errors, which are not plotted here. We see a number of 2D-candidates above the isochrone at different absolute magnitudes that correspond to different masses. The WW-candidates extend to much lower magnitudes than the RW-candidates shown in Fig.~\ref{fig:2D_RW_candidates_after_correction}. On the CMD, we identify the candidates we have in common with \citet{2019ApJ...884....6M} (green square), which sit towards brighter magnitudes but still within the pre-main sequence part of the isochrone. Many of our identified candidates have large errors in magnitude and colour, which are predominantly driven by the errors in the extinction and reddening. Quite a few of these candidates sit below the 4 Myr isochrone, but might be younger than their position suggests due to the large errors.}
        \label{fig:2D_WW_candidates_after_correction}
\end{figure*}

\begin{table*}
    \renewcommand\arraystretch{1.045}
	\centering
	\caption{RW star 2D candidates sorted by decreasing 2D-velocity. Column 2+3: velocity in ONC rest frame [rf]; Column 3: RV sources - $^{a}$\textit{Gaia} DR2, $^{b}$\citet{2015ApJ...807...27C}; Column 4: indication of 3D-candidate status; Column 5: minimum flight time since ejection (crossing of search boundary); Column 6: age from PARSEC isochrones \citep{RN225}; Column 7--9: from literature sources - $^{1}$\citet{1988AJ.....95.1744V}, $^{2}$\citet{1997AJ....113.1733H}, $^{3}$\citet{1999MSS...C05....0H}, $^{4}$\citet{2000AJ....119.3026R}, $^{5}$\citet{2016ApJ...818...59D}, $^{6}$\citet{1995A&AS..110..367N}.}
	\label{tab:RWC_2D}
	\begin{tabular}{lcccccccr} % four columns, alignment for each
		\hline
		\textit{Gaia} DR2 source-id & 2D-velocity rf & Radial velocity rf & 3D-cand. & Flight time & Iso. age & Age & Mass & Spectral type  \\
		 & (km\,s$^{-1}$) & (km\,s$^{-1}$) & & (Myr) & (Myr) & (Myr) & (M$_{\sun}$) \\
		\hline
		3216203177762381952	&104.1	$\pm$0.4 &	65.6 $\pm$12.8$^{a}$	&	yes	&	0.2	&0.4\,$^{+	9.0	}_{	-0.3}$	&-&-&-\\
        3012378357905079040	&98.3	$\pm$0.4	&	-			&	-	&	0.4	&	10.0 $\pm$6.0&	-	&-&-\\	
        3015321754828860928	&82.9   $\pm$1.0   &   -           &   -   &   0.3 &  0.4\,$^{+1.1}_{-0.3}$ & - & - & -\\
        3013484917577226240	&68.8$\pm$0.4	&	-			&	-	&	0.6	&   3.0\,$^{+5.0}_{	-2.0}$	&	-	&-&-\\	
        3012438796685305728	&59.9	$\pm$0.5	&	0.0	$\pm$6.8$^{a}$	&	yes	&	0.5	&	10.0\,$^{+	40.0}_{	-7.5}$	&	-	&-&-\\	
        2986587942582891264	&59.8	$\pm$0.5	&	-21.3 $\pm$6.6$^{a}$	&	yes	&	1.2 &	9.0\,$^{+	10.0}_{	-6.0}$	&	-	&- &-\\	
        3016780428803888768	&55.4	$\pm$0.3	&	-			&	-	&	0.2	&	1.0\,$^{+5.0}_{	-0.7}$	&	-	&	-&-	\\	
        3012393858443805440	&52.9	$\pm$0.4	&	-			&	-	&	0.7	&	12.0\,$^{+35.0}_{-8.0}$	&	-	&-&	-\\
        3209498394512739968	&47.2	$\pm$0.4	&	-			&	-	&	0.1	&	2.0	$\pm$1.0&	-	&-&M3$^{4}$\\
        2995865656058327552	&46.6	$\pm$0.4	&	-			&	-	&	1.3	&	0.1\,$^{+1.9}_{	-0.1}$	&	-&-&-\\	
        2989899774685582592 &46.2   $\pm$0.4	&	3.5  $\pm$6.6$^{a}$ & no & -& 0.1\,$\pm$0.1 & - & - & -\\
        3016792935748254336	&46.0	$\pm$0.5	&	-			&	-	&	0.2	&   0.3	$\pm$0.2&	-	&-&-\\	
        3122639449820663040	&43.9	$\pm$0.5	&	-			&	-	&	1.6	&	2.0\,$^{+2.0}_{	-0.5}$	&	-&-& B5$^{3}$\\
        3320554665258533376	&37.0	$\pm$0.5	&	-			&	-	&	2.0	&	5.0	$\pm$2.0&	-	&-&	-\\
        2998537847270106240	&36.3	$\pm$0.4	&	45.3 $\pm$6.6$^{a}$	&	no	& -	& 0.1\,$^{+	0.1	}_{	-0.1}$ &-&-&G8$^{3}$\\
        2998697894931641600	&36.0	$\pm$0.5	&	-			&	-	&	1.2	&	2.0	$\pm$0.2&	-&-& A2-A9$^{3}$\\
        3003060825792025088	&35.9	$\pm$0.4	&	36.9 $\pm$6.7$^{a}$	&	no	&	- &	8.0	$\pm$5.0	&-&-&-\\
        3023329085698084992	&34.2	$\pm$0.5	&	-			&	-	&	0.2	&	1.0\,$^{+6.0}_{	-0.9}$&	-	&-&	-\\
        3209936343738052992	&32.9	$\pm$0.4	&	-			&	-	&	0.3	&	1.5\,$^{+8.5}_{	-1.0}$	&-&-&-\\
        3017250019053914368 &31.9   $\pm$0.4    & 32.5 $\pm$6.6$^{a}$ & yes & in cluster & 8.0\,$^{+42.0}_{	-7.0}$ & 1.9-5.7$^{2,5}$ & 1.9-2.4$^{2,5}$ & G6$^{2}$    \\
        3122561556293863552	&30.3	$\pm$0.4	&	-17.5 $\pm$10.7$^{a}$	&	yes	&	1.8	& 1.8\,$^{+	2.1}_{	-1.1}$ & - & - & -  \\
        3017265515291765760	&30.1	$\pm$0.4 &	12.2 $\pm$6.6$^{a}$	&	yes	&	in cluster	&	2.2\,$^{+4.8}_{	-2.0}$&	0.3$^{2}$	&	2.5$^{2}$	&	K1$^{1}$\\
        3218162816720967040	&30.0	$\pm$1.1	&	-	&	-	&	1.4	& 15.0\,$^{+	40.0}_{	-11.2}$ & - & - & -  \\
        3222673430030590592	&29.5   $\pm$0.4   &   7.4 $\pm$6.6$^{a}$    & yes* & 1.5* & 0.3\,$\pm$0.2 & - & - & K0$^{6}$\\ 
        3016016714897329152	&29.0	$\pm$0.3	&	-27.7 $\pm$6.6$^{a}$	&	no	&	-	& 7.0	$\pm$4.0 & - & - & -  \\
        3016354436766366848 &28.1   $\pm$0.4    &  **$^{b}$   &   ? &   0.4 & 4.0 $\pm$1.0  & - &-& A0$^{3}$\\
        2988494014709554176	&27.4	$\pm$0.3	&	-24.3 $\pm$7.1$^{a}$	&	yes	&	2.3	& 10.0\,$^{+35.0}_{	-6.0}$ & - & - & - \\
        3208970285334738944	&22.7	$\pm$0.4	&	-29.9 $\pm$18.1$^{a}$	&	yes	&	0.5 & 2.8\,$^{+	8.2}_{	-2.4}$ & - & - & -  \\
        3015532208227085824	&21.8	$\pm$0.4	&	-54.5 $\pm$6.7$^{a}$	&	yes	&	0.8	& 6.5	$\pm$3.5 & - & - & - \\
        3216868764551493504	&20.4	$\pm$0.4	&	45.2 $\pm$6.8$^{a}$ 	&	no	&	-	& 5.0\,$^{+	40.0}_{	-4.5}$ & - & - & -  \\
        3122421987035894784	&20.0	$\pm$0.4	&	52.7 $\pm$6.6$^{a}$	&	no	&	- & 0.4\,$^{+	0.6}_{	-0.2}$ & - & - & -  \\
		\hline
		\multicolumn{9}{l}{\parbox[t]{17cm}{*age estimate is smaller than the flight time; **multiple and varying RV measurements, possibly indicating a binary system}}\\
	\end{tabular}
\end{table*}

\begin{table*}
    \renewcommand\arraystretch{1.045}
	\centering
	\caption{WW star 2D candidates. Column 2+3: velocity in ONC rest frame [rf]; Column 3: RV sources - $^{a}$\textit{Gaia} DR2, $^{b}$\citet{2015ApJ...807...27C}, $^{c}$\citet{2006AstL...32..759G},$^{d}$\citet{2018AJ....156...84K}; Column 4: indication of 3D-candidate status; Column 5: minimum flight time since ejection (crossing of search boundary); Column 6: age from PARSEC isochrones \citep{RN225}; Column 7--9: from literature sources - $^{1}$\citet{1988AJ.....95.1744V}, $^{2}$\citet{1997AJ....113.1733H}, $^{3}$\citet{1999MSS...C05....0H}, $^{4}$ \citet{2000AJ....119.3026R}, $^{5}$\citet{RN218}, $^{6}$\citet{RN263}, $^{7}$\citet{2012ApJ...748...14D}, $^{8}$\citet{2012ApJ...752...59H}, $^{9}$\citet{2013ApJ...764..114H}, $^{10}$\citet{2016ApJ...818...59D}, $^{11}$\citet{2011ApJS..193...24S}, $^{12}$\citet{2010AN....331..349H}.}
	\label{tab:WWC_2D}
	\begin{tabular}{lcccccccc} % four columns, alignment for each
		\hline
		\textit{Gaia} DR2 source-id & 2D-velocity rf & Radial velocity rf & 3D-cand. & Flight time & Iso. age & Age & Mass & Spectral type  \\
		 & (km\,s$^{-1}$) & (km\,s$^{-1}$) & & (Myr) & (Myr) & (Myr) & (M$_{\sun}$) & \\
		\hline
		3222368036380921600	&26.7	$\pm$0.4 &	-8.9 $\pm$13.7$^{a}$&	yes	&	1.3	& 3.0\,$^{+6.0}_{-2.0}$	& - & - & -\\
        3209653627514662528 &25.8	$\pm$0.5	&   12.9 $\pm$6.6$^{b}$	& no  &- &	0.4	$\pm$0.1& 0.2$^{10}$	&	0.3$^{10}$	&	-	\\
        3013902388397518208	&25.7	$\pm$0.4 &		-		&	-	&	1.0		&	1.0$\pm$0.5	&	-	&	-	&	-	\\	
        3209590577396377856	&25.2	$\pm$1.1	&		-		&	-	&	0.1		&	5.0\,$^{+15.0}_{-4.0}$	&	-	&	-	&	M6$^{4}$	\\	
        3231583219428074752	&25.1	$\pm$0.3	&	-	&	-	&	2.9		& 8.0\,$^{+4.0}_{-5.0}$	& -&-&-\\
        3209228906787768832	&24.8	$\pm$0.4	&	-8.4 $\pm$6.8$^{a}$	&	yes	&	0.8	& 10.0\,$^{+	41.0}_{	-8.3}$ & - & - & -  \\
        3004263966389331456	&22.4	$\pm$0.4	&	-15.4 $\pm$6.6$^{a}$	&	yes	&	3.0	& 6.0\,$\pm$2.5 & - & - & -  \\
        3181732702253990144	&22.2	$\pm$0.4	&	-	&	-	&	2.9	& 1.0\,$^{+	2.0}_{	-0.5}$ & - & - & B9$^{3}$  \\
        3017260292611534848 &21.7  $\pm$0.4    &   3.5 $\pm$6.7$^{a}$ & yes & 0.1 &  3.0\,$^{+7.0}_{	-1.2}$ & 7.7$^{10}$ & 1.9$^{10}$ & -  \\
        3023589704313257600	&19.7	$\pm$0.6	&		-		&	-	&	1.0*	& 0.2\,$^{+0.4}_{-0.1}$	&	-	&	-	&	-	\\	
        3023540054490074752	&18.6	$\pm$1.7	&		-		&	-	&	0.6		&	10.0\,$^{+20.0}_{-8.0	}$	&	-	&	-	&	-	\\
        3015018014743100544	&18.5	$\pm$0.4	&	-21.2 $\pm$20.7$^{a}$	&	yes	&	1.4 &	0.9\,$^{+3.1}_{-0.6}$&	-	&-	&	-\\
        3014981937018182144	&18.4	$\pm$0.7	&		-		&	-	&	1.7		&	1.2\,$^{+2.9}_{-0.7	}$	&	-	&	-	&	-	\\	
        3209624872711454976 &18.1	$\pm$0.4	&	14.7 $\pm$6.6$^{b}$	&	yes	&	0.2	&	1.7	$\pm$0.6&	0.4$^{10}$&	0.5$^{10}$	&	-\\	
        3209836391259368960	&17.4	$\pm$1.3	&		-		&	-	&	0.3		&	10.0\,$^{+20.0}_{-8.0}$	&	-	&	-	&	-	\\
        3017166907140904320 &17.2   $\pm$0.4    &   5.3 $\pm$6.6$^{b}$  & yes & 0.2 & 2.8\,$^{+0.3}_{-0.9}$ & 1.0$^{10}$ & 0.6$^{10}$ & K7.5$^{8}$  \\
        3017242051888552704	&16.7	$\pm$0.4	&	-5.5 $\pm$6.9$^{b}$	&	yes	&	in cluster	&	4.0\,$^{+	20.0	}_{	-3.5	}$	&	1.8$^{10}$	&	0.7$^{10}$	& -		\\
        3209497088842680704 &16.5	$\pm$0.4	&	-	&	-	&	0.2		&	3.0	$\pm$1.5 &	-	&	-	&	M2$^{4}$\\	
        3015625563635553024	&16.5	$\pm$0.4	&	-3.5 $\pm$6.6$^{b}$	&	yes	&	0.9		&	0.8\,$^{+	3.2	}_{	-0.4	}$ 	&	1.2$^{10}$ & 0.3$^{10}$&	M2.9$^{8}$	\\	
        3209424108758593408 &16.4	$\pm$0.4	&	8.3 $\pm$6.6$^{b}$	&	yes	&	in	cluster	&	1.0\,$^{+9.0}_{-0.9}$&0.5-2.5$^{5,11}$	&1.1-2.3$^{5,11}$&G9$^{8}$	\\
        3016070590967059968	&16.4	$\pm$0.6	&	-4.5 $\pm$6.6$^{d}$		&	yes	&	1.0	&	0.5\,$^{+	1.5	}_{	-0.4	}$	&	-	&	-	&	-	\\
        3214878167468186880 & 15.9 $\pm$0.5	& - & - & 2.4 & 2.1\,$\pm$1.1 & - & - & - \\
        3220151695816273152	&15.6	$\pm$0.4	&	10.0	$\pm$6.6$^{a}$	&	yes	&	2.2	&	2.1\,$^{+	3.0	}_{	-1.5	}$ 	&	-	&	-	&	-	\\
        3017402614955763200	&14.8	$\pm$0.4	&	-13.7 $\pm$8.1$^{a}$&yes	&	0.1	&	6.0	$\pm$4.0	&	-	&	-	&	K7$^{4}$	\\	
        3208349129984108800 & 14.8  $\pm$0.4    & - & -  & 2.4 & 2.1\,$\pm$1.1 & - & - & - \\  
        2984454031031531008 &14.8  $\pm$0.4    & - & -     & 3.8 &  2.8\,$^{+3.2}_{-1.3}$ & - & - & - \\
        3209424108758593536 &14.1	$\pm$0.4	&	-4.3 $\pm$6.6$^{b}$	&	yes	&	in	cluster	&	4.0\,$^{+	6.0}_{-3.3	}$	&	0.5-2.5$^{5,11}$	&	0.7$^{5,11}$&	K7$^{2}$\\
        3017384129418196992	&14.1	$\pm$0.4	&		-		&	-	&	0.1		&	2.5\,$^{+	1.5	}_{	-0.5	}$	&	-	&	-	&	M2$^{4}$	\\	
        2984723926777044480 &13.9   $\pm$0.4&   -10.1 $\pm$6.6$^{a}$ &yes* & 3.4* & 0.4$\pm$0.3 & - & -\\
        3012142379518284288	&13.5   $\pm$0.4&		-		&	-	&	2.0	& 2.1\,$^{+	3.0	}_{	-1.1	}$	&	-	&	-	&	-	\\			
        3015334914608642688	&13.3	$\pm$0.5	&	17.3 $\pm$7.6$^{b}$	&	yes	&	1.8	&	1.5\,$^{+	3.5	}_{	-1.0}$	&	-	&	-	&	M1.6$^{8}$	\\
        3209074803362165888	&13.2	$\pm$0.4&		-		&	-	&	0.6	&	0.8\,$^{+	9.0	}_{	-0.7	}$	&	-	&	-	&	-	\\
        3014834946056441984	&13.1	$\pm$0.4&		-		&	-	&	1.6	&	1.0\,$^{+	0.8	}_{	-0.4	}$	&	1$^{6}$	&	3.8$^{6}$&	A5$^{6}$	\\
        3209531650444835840 &13.0	$\pm$0.4	&	-4.2 $\pm$7.8$^{a}$	&	yes	&	in	cluster	&	0.3\,$^{+	3.7	}_{	-0.2	}$ 	&	-	&	3.8$^{2}$	&	K0$^{1}$	\\
        3012280432650658304	&12.9	$\pm$3.6	&	-	&	-	&	2.9	& 14.0\,$^{+	36.0}_{	-10.5}$ & - & - & -  \\
        3216889827071056896	&12.6	$\pm$0.6	&		-		&	-	&	1.4	&	1.0\,$^{+	3.0	}_{	-0.9	}$	&	-	&	-	&	-	\\	
        3013899158582179712	&12.4	$\pm$1.4	&	-	&	-	&	1.9	& 5.0\,$^{+	7.0	}_{	-4.0}$ & - & - & -  \\
        3216174629116142336	&12.1	$\pm$0.4	&		-		&	-	&	1.7	&	2.5	$\pm$1.5&	-	&	-	&	-	\\
        3016101579155228928	&12.1   $\pm$1.4	&	-	&	-	&	1.1	& 1.0\,$^{+	10.0	}_{	-0.8}$ & - & - & -  \\
        3219378365481960832 &11.7   $\pm$0.4	&	-	&	-	&	3.3		& 2.8\,$^{+	7.0}_{	-1.8}$ & - & - & -\\
        3017341385903759744	&11.7	$\pm$0.4	&	1.1	$\pm$6.6$^{d}$	&	yes	&	in	cluster	&	3.0$\pm$1.0	&	0.7-0.9$^{5,11}$ &	0.5$^{5,11}$ &	K6$^{4}$\\
        3017340664349130368	&11.3	$\pm$0.4	&	14.7 $\pm$16.9$^{a}$	&	yes	&	0.3		&	2.0\,$^{+	10.0	}_{	-1.7	}$&	-	&	-	&	K5$^{4}$	\\	
        3209527291054667136	&11.1	$\pm$0.5	&	10.6	$\pm$6.6$^{b}$	&	no	&	-	&	1.2\,$^{+	1.8	}_{	-1.1	}$&	0.8-2$^{5,11}$	&2.5-3.2$^{5,11}$	&-	\\
        3017260022031719040	&10.6	$\pm$0.4	&	8.7	$\pm$6.6$^{d}$	&	yes	&	in  cluster	& 2.5	$\pm$1.5	&	0.7-1.5$^{5,7}$& 0.3-0.5$^{5,7}$&	M3$^{5}$	\\
        3017270879709003520	&10.5	$\pm$0.4	&	-1.2 $\pm$6.8$^{b}$	&	yes	&	in	cluster	&	0.9\,$^{+	5.0	}_{	-0.8	}$	&	0.3$^{10}$	&	1.4$^{10}$	&	K4$^{1}$\\
        3016961676421884672 &10.4   $\pm$0.6    & - & -     & 0.8 &	0.6\,$^{+	1.3	}_{	-0.5	}$ & - & - & - \\
        3016424530632449280 &10.2 $\pm$1.5 &  -4.9 $\pm$7.1$^{c}$& no & - & $\sim$4 & - & 24.1$^{12}$ & O9.7$^{11}$\\
        3215804677813294976 & 10.1 $\pm$0.8    & - & - & 1.2 & 0.5\,$^{+	1.5}_{	-0.1}$ & - & - & - \\ 
        3017252600328207104 & 9.3  $\pm$0.4    & -4.9$\pm$6.6$^{d}$ & yes & 0.1 &  1.2\,$^{+	2.8}_{	-0.5}$	& 0.1$^{10}$ & 0.3$^{10}$ & M3$^{2}$ \\
	    3209680466766984448 & 9.1   $\pm$0.4	&	16.0	$\pm$6.6$^{a}$	&	yes	&	1.4 & 0.4\,$^{+	1.5}_{	-0.3}$	& - & - & - \\
        3017174741161205760	& 8.7	$\pm$0.4	&	-17.2	$\pm$9.1$^{a}$	&	yes	&	0.3		& 4.0\,$^{+	20.0	}_{	-3.5	}$	& - & - & K3$^{9}$\\
        3217017610938439552 & 6.3    $\pm$0.4	&	-8.2 $\pm$11.1$^{a}$	&	yes	&	3.2	& 1.5\,$^{+	6.5}_{	-1.2}$ & - & - & -  \\
        3017367151399567872	&	3.6	$\pm$0.4	&	12.2	$\pm$9.0$^{a}$	&	yes	&	in cluster	& 4.0	$\pm$2.0 		&	1.7$^{10}$	&	2.7$^{10}$	&-	\\
        3209529112120792320	&	3.0	$\pm$0.4	&	12.1	$\pm$6.6$^{b}$	&	yes	&	in cluster	& 20.0\,$^{+	50.0	}_{-19.0	}$ &	6.2$^{10}$	&	1.1$^{10}$	&	-	
        \\
        \hline
        \multicolumn{9}{l}{\parbox[t]{17cm}{*age estimate is smaller than the flight time}}
	\end{tabular}
\end{table*}

\begin{figure*}
    \centering
	\includegraphics[width=0.95\linewidth]{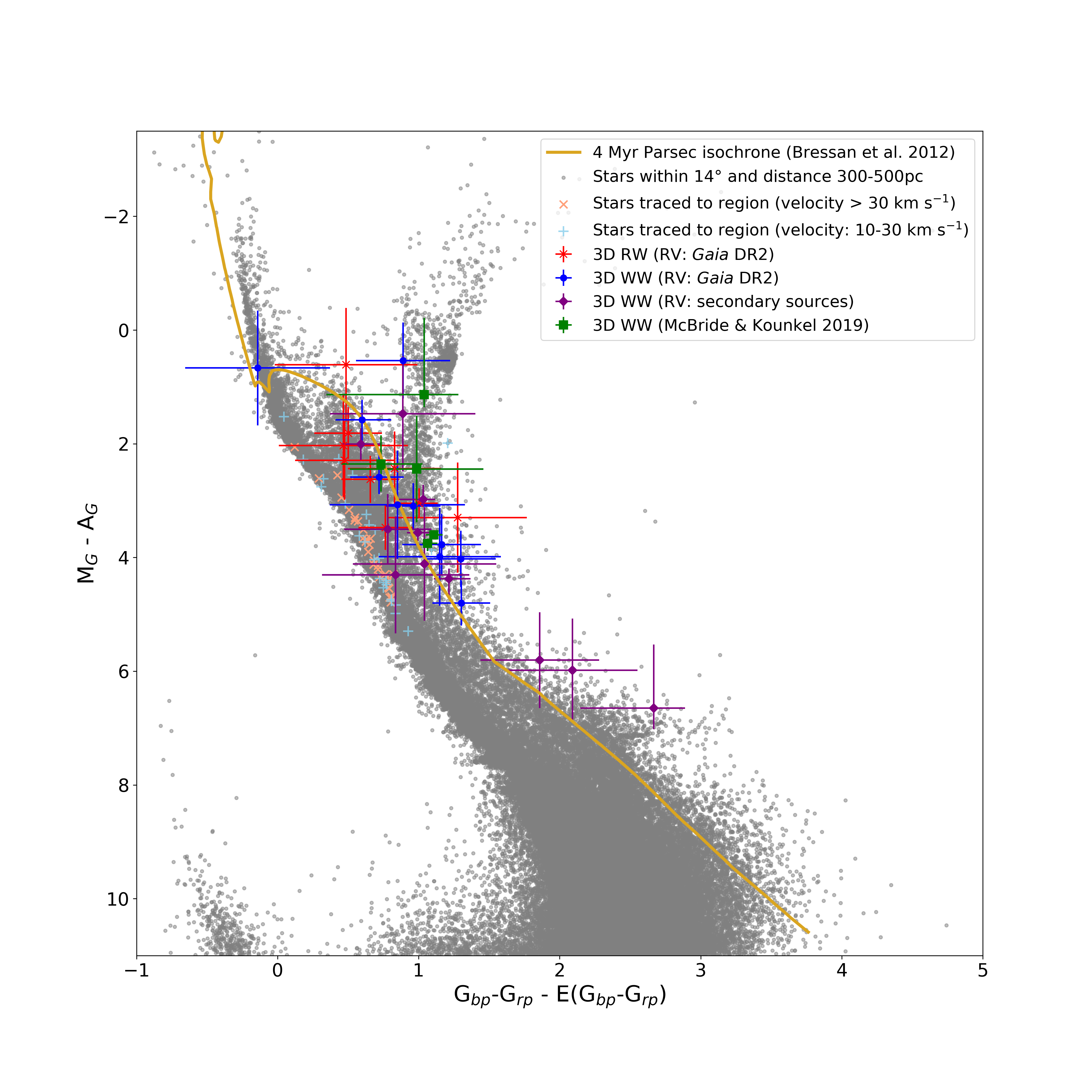}
    \caption{Colour magnitude diagram showing all 3D-candidates (at RW and WW velocities). We magnitude limit the diagram to -3.5\,mag < $M_{\rm{G}}-A_{\rm{G}}$ < 11\,mag, which corresponds to a $G$-magnitude $\approx$ 19\,mag at the fainter end. Around this apparent magnitude value, the typical uncertainties in the 5-parameter astrometry increase quickly, however our faintest 3D-candidate is much brighter than this limit and has an absolute magnitude of $\sim$7\,mag. 
    Our confirmed 3D-candidates are all located towards the upper end of the 4 Myr isochrone. All of our 3D RW stars (red ``x'') are identified using \textit{Gaia} DR2 RV. Among the 3D WW stars we separately show the candidates with RV from \textit{Gaia} DR2 (blue square) and from secondary sources \citep{2015ApJ...807...27C,2018AJ....156...84K} (purple dot). Finally we show the five 3D WW stars what we have in common with \citet{2019ApJ...884....6M} (green square). Many of our identified candidates have large errors in magnitude and colour, which are predominantly driven by the errors in the extinction and reddening. Interestingly, we also find a small number of 3D trace-backs that sit along the MS, below the isochrone (with RW-velocity: orange ``x'', with WW-velocity: light-blue ``+'') and even when considering their errors they are too old to have been born in the ONC. These might be past visitors to the ONC or could possibly belong to an older population of stars, as their estimated isochronal ages are between 5--40 Myr.}
    \label{fig:3D-candidates_after_correction_new}
\end{figure*}

\section{Results from \textit{Gaia} DR2}

\subsection{2D-candidates}

Fig.~\ref{fig:2D_RW_candidates_after_correction} and \ref{fig:2D_WW_candidates_after_correction} show the resulting CMDs following the procedure described in Sec.~\ref{CMD_constr} for our 2D-candidate RW and WW stars. A large number of the stars that have been traced back to the ONC search region in the xy-plane (on the sky) are located along the MS underneath the 4 Myr isochrone and are too old to have originated from the ONC. Located above the isochrone are all traced-back candidates that are young enough to have been born in the ONC. In addition, we also find potential candidates that are located below the isochrone but where the photometric error bars cross the isochrone, indicating a possibly younger age.

We find 31 RW and 54 WW 2D candidates with an isochronal age < 4 Myr after applying our criteria for candidate identification. We exclude 1 RW and 1 WW candidate without RV as based on their radial distance to the ONC the RV required to get to their current position since ejection from the ONC is unreasonably large for runaway or walkaway stars (> |500|\,km\,s$^{-1}$). A few 2D-candidates (7 RWs and 8 WWs) with known RVs are excluded as their RVs point towards the ONC instead of away from it, so these stars cannot have originated from the ONC.

Table~\ref{tab:RWC_2D} provides an overview of all our identified RW candidates in 2D with information about their velocities in the ONC rest frame, flight times and approximate isochronal ages. We also identify whether any of the stars with RV have been traced back successfully in 3D. Table~\ref{tab:WWC_2D} provides the same information for the WW candidates. Both tables also include information gathered from literature sources about our candidates, such as age, mass and spectral type. In Appendix~\ref{App_excluded_cand}, Table~\ref{tab:RWC_2D_excluded} lists the excluded 2D-candidates, including their reason for being excluded.

The brightest 2D RW candidate HD 288089 (\textit{Gaia} DR2 3222673430030590592) has an absolute magnitude of $\sim$-0.2\,mag. Very little is known about this star, apart from its spectral type listed as K0 \citep{1995A&AS..110..367N}. It is the only 2D RW candidate whose flight time is larger than its isochronal age and as a consequence it is unlikely to have originated from the ONC. The second brightest star in this list HD 41288 (\textit{Gaia} DR2 3122639449820663040) has an absolute magnitude of $\sim$0\,mag and its spectral type from literature is a B5 \citep{1999MSS...C05....0H}. This candidate is the most massive 2D RW candidate in our data set and its position is consistent with an isochrone age of $\sim$2 Myr. Its 2D-velocity is $\sim$44\,km\,s$^{-1}$.

Our brightest 2D WW candidate $\upsilon$~Ori (\textit{Gaia} DR2 3016424530632449280) is a known O-star \citep{2011ApJS..193...24S} with an absolute magnitude of $\sim$-3.3\,mag on our CMD. This star is one of only a few stars in our 2D-candidate list that has reached the MS. It is located slightly underneath the MS due to an over-correction for extinction as a consequence of our chosen approach. Its mass is reported to be $\sim$24 M$_{\sun}$ \citep{2010AN....331..349H}. It is located at the very edge of our search field with a distance of just under 100 pc to the ONC centre. 

Several of our faintest 2D candidates appear not to have any further information than that contained within \textit{Gaia} DR2. The faintest candidates will also be the stars that have the lowest mass, so information from other sources is critical. The faintest 2D RW candidate with a spectral type identification is V* HP Ori (\textit{Gaia} DR2 3209498394512739968). \citet{2000AJ....119.3026R} suggested it to be a M3-type star. Its absolute magnitude is $\sim$3.5\,mag. The faintest 2D WW candidate 2MASS J05332200-0458321 \textit{Gaia} DR2 3209590577396377856) has an absolute magnitude of $\sim$10\,mag and a suggested spectral type of M6 \citep{2000AJ....119.3026R}.

\subsubsection{Comparison to McBride \& Kounkel 2019}
\citet{2019ApJ...884....6M} identified 9 ``runaway stars'' with an origin close to the Trapezium cluster at the centre of the ONC. Of these stars, we successfully trace back 7 in our analysis, however none fit our velocity requirement of a RW (velocity > 30\,km\,s$^{-1}$) and all are identified as WW candidates instead:
\begin{itemize}
    \item V1961 Ori (\textit{Gaia} DR2 3209424108758593408): this star has previously been identified as a ``runaway'' candidate by \citet{2017ApJ...834..142K}. It has been suggested as a spectroscopic binary by \citet{2019AJ....157..196K}, but has a RUWE $\approx$ 1.1 in \textit{Gaia} DR2 indicating a good fit to a single-star model or that this system's binary status does not affect its astrometric quality. It has the second lowest $M_{\rm{G}}-A_{\rm{G}}$ value of $\sim$3.7\,mag of all \citet{2019ApJ...884....6M} stars found in our analysis, but as a binary will be brighter than each of the individual stars. Its 2D-velocity in the ONC rest frame is $\sim$16\,km\,s$^{-1}$.
    \item Brun 259 (\textit{Gaia} DR2 3209424108758593536): \citet{RN257} show it is unlikely to be binary system. It has a rest-frame 2D-velocity of $\sim$15\,km\,s$^{-1}$ and an absolute magnitude of $\sim$2.4\,mag.
    \item V1321 Ori (\textit{Gaia} DR2 3209531650444835840): \citet{2012ApJ...754...44J} suggest this star is possibly a binary, its RUWE $\approx$ 1 indicates a good fit to the single star model. This low value indicates that the binarity of this system is not affecting the astrometry quality. It has a 2D-velocity of $\sim$14km\,s$^{-1}$ and an absolute magnitude $\sim$1.1\,mag.
    \item V1440 Ori (\textit{Gaia} DR2 3209624872711454976): this star has an absolute magnitude $\sim$1.1\,mag and a 2D-velocity of of $\sim$18\,km\,s$^{-1}$ in the ONC rest frame.
    \item 2MASS J05351295-0417499 (\textit{Gaia} DR2 3209653627514662528): this star is one of our fastest 2D-candidates with a 2D-velocity of $\sim$26\,km\,s$^{-1}$. It is one of the youngest in our WW list with an isochronal age of $\sim$0.4 Myr and a magnitude of $\sim$3.4\,mag.
    \item CRTS J053223.9-050523 (\textit{Gaia}~DR2 3209497088842680704): the 2D-velocity of this star is $\sim$17\,km\,s$^{-1}$ and it has a magnitude of $\sim$2.9\,mag. 
    \item Haro 4-379 (\textit{Gaia} DR2 3017166907140904320): this star is the faintest in this \citet{2019ApJ...884....6M} group with a $M_{\rm{G}}-A_{\rm{G}}$ value of $\sim$3.8\,mag and a 2D-velocity of $\sim$17\,km\,s$^{-1}$.
\end{itemize}
The final two candidates of \citet{2019ApJ...884....6M} have been excluded from our data set from the outset due to their high RUWE-value, which indicates that the astrometry values might be unreliable. 2MASS J05382070-0610007 (\textit{Gaia} DR2 3016971567730386432) has a RUWE $\approx$ 8 and V360 Ori (\textit{Gaia} DR2 3209528081326372864) has a RUWE $\approx$28, this latter star is also a known binary \citep{2012A&A...540A..46D}.

\subsection{3D-candidates}
Fig.~\ref{fig:3D-candidates_after_correction_new} shows the CMD of RW and WW stars that can be traced back in 3D to our search region. Ten of the 31 RW 2D-candidates are successfully traced back in all three dimensions using \textit{Gaia} DR2 RV.

The fastest RW star (\textit{Gaia} DR23216203177762381952) in our sample has an ONC rest frame space velocity of $\sim$123\,km\,s$^{-1}$ and has an absolute magnitude of $\sim$4.4\,mag, however there is no additional information available about this star.  
Two of the 10 3D RWs have information available from the literature about their spectral type, mass and/or age. Brun 609 (\textit{Gaia} DR2 3017250019053914368) has a G6 spectral type \citep{1988AJ.....95.1744V}, a mass estimate of $\sim$1.9--2.4 M$_{\sun}$ and an age estimate of 1.9--5.7 Myr \citep{1997AJ....113.1733H,2016ApJ...818...59D}. It has a rest frame space velocity of $\sim$46\,km\,s$^{-1}$ and is still located within the search boundary.

BD-05 1307 (\textit{Gaia} DR2 3017265515291765760) is a K1-type star \citep{1988AJ.....95.1744V} with a mass estimate of $\sim$2.5 M$_{\sun}$ and an estimated age of $\sim$0.3 Myr \citep{1997AJ....113.1733H}. It has a rest frame space velocity of $\sim$33\,km\,s$^{-1}$. It is also the brightest 3D RW star in our data set with an absolute magnitude of $\sim$0.7\,mag and is still located within the central 2.5 pc search boundary at a distance of $\sim$400 pc.

We find another potential 3D RW star TYC 4774-868-1 (\textit{Gaia} DR2 3209532135777678208). This star could be an example of a special case described in \citet{RN309}. These authors found stars that appear bound in proper motion or 2D-velocity space and are still located in the cluster, however a very high RV turns these stars into RWs. If we only consider its 2D-velocity of $\sim$3.4\,km\,s$^{-1}$, TYC 4774-868-1 appears to be still bound to the ONC. It is also still located in the central region. However its \textit{Gaia} DR2 RV in the ONC rest frame would be high enough to turn it into a RW star. There is a caveat as \citet{2015ApJ...807...27C} states a much lower RV of $\sim$28.3 km/s, which results in a RV of $\sim$6.5 km/s in the ONC rest frame. This leads us to not consider this star as a RW until further clarification of its RV. The data for this star is contained in Table~\ref{tab:2D_cand_varying_RV} in Appendix~\ref{Varying_RV}.

The brightest 2D RW candidate HD 288089 (\textit{Gaia} 3222673430030590592) also traces back in three dimensions, however its isochronal age is smaller than its flight time. While it traces back to the ONC, it cannot have been born there, but must have instead come from somewhere in-between the ONC and its current position. It is not counted in the final list of 3D RWs and this leaves us with 9 3D RWs.

Of the 54 WW 2D-candidates, 27 are also 3D-candidates. Fourteen of these using \textit{Gaia} DR2 RV, another 9 candidates using RV from \citet{2015ApJ...807...27C} and 4 candidate using RV from \citet{2018AJ....156...84K}. Five of the 7 2D WW candidates we have in common with \citet{2019ApJ...884....6M} are also 3D WW candidates:
\begin{itemize}
    \item V* V1961 Ori \textit{Gaia} DR2 3209424108758593408): has a G9 spectral type \citep{2012ApJ...752...59H} with an estimated mass of $\sim$1.1--2.3~M$_{\sun}$ and an age of $\sim$0.1--2.5 Myr \citep{RN218, 2016ApJ...818...59D}. Its space velocity in our rest-frame is $\sim$18\,km\,s$^{-1}$.
    \item Brun 259 \textit{Gaia} DR2 3209424108758593536): has a spectral type K7 \citep{1997AJ....113.1733H} with a mass of $\sim$0.7 M$_{\sun}$ and an age estimate of $\sim$0.5--2.5 Myr \citep{RN218,2016ApJ...818...59D}. It has a rest-frame space velocity of $\sim$15\,km\,s$^{-1}$.
    \item V1321 Ori \textit{Gaia} DR2 3209531650444835840): this star has a spectral type K0 \citep{1988AJ.....95.1744V} and a mass estimate of 3.8 M$_{\sun}$ \citep{1997AJ....113.1733H}. Its space velocity is $\sim$14\,km\,s$^{-1}$.
    \item V1440 Ori \textit{Gaia} DR2 3209624872711454976) is one of the faster 3D WW stars with an ONC rest frame space velocity of $\sim$23\,km\,s$^{-1}$, it is a sub-solar mass ($\sim$0.5 M$_{\sun}$) star with an age estimate of $\sim$0.4 Myr \citep{2016ApJ...818...59D}, suggesting it left the cluster shortly after its birth.
    \item Haro 4-379 \textit{Gaia} DR2 3017166907140904320): has a space velocity $\sim$18\,km\,s$^{-1}$ with a spectral type of K7.5 \citep{2012ApJ...752...59H}, has a sub-solar mass ($\sim$0.6 M$_{\sun}$) and an age estimate from literature of $\sim$1 Myr \citep{2016ApJ...818...59D}.
\end{itemize}

\noindent The slowest confirmed 3D WW star is Brun 519 (\textit{Gaia} DR2 3017252600328207104) and has a rest frame space velocity just above the 10\,km\,s$^{-1}$ lower velocity limit. It is also one of the youngest 3D WW star in our sample with an age and mass estimate of $\sim$0.1 Myr and $\sim$0.3 M$_{\sun}$, respectively \citep{2016ApJ...818...59D}. Its spectral type is suggested to be a M3 \citep{1997AJ....113.1733H}.

Like we have for the 3D RW candidates, we also find one 3D WW candidate BD-13 1169 (\textit{Gaia} DR2 2984723926777044480), where the flight time since ejection is considerably larger than the estimated isochronal age, strongly indicating that this star was not born in the ONC. We exclude this star from the 3D WW list and are left with 26 3D WW stars.

We also find three further potential 3D WW stars that we might still consider to be bound to the ONC if we only had information about their proper motion (or 2D-velocities). Here a larger RV (from \textit{Gaia} DR2 or secondary sources) can turn these stars into 3D candidates. These WW candidates (like the RW candidate TYC 4774-868-1) show very different RV measurements in different surveys, which could have an alternative explanation of a bound binary system. These candidates are shown separately in Table~\ref{tab:2D_cand_varying_RV} in Appendix~\ref{Varying_RV}.

We also find several older 3D RW/WW stars below the isochrone in Fig.~\ref{fig:3D-candidates_after_correction_new} along the main sequence. These stars are possibly past ``visitors'' to the ONC, having travelled through the ONC from their origin. Information about these visitors is provided in Appendix~\ref{App_visitors} in Table~\ref{tab:Visitors_list}.

\begin{figure*}
        \centering
        \includegraphics[width=0.95\linewidth]{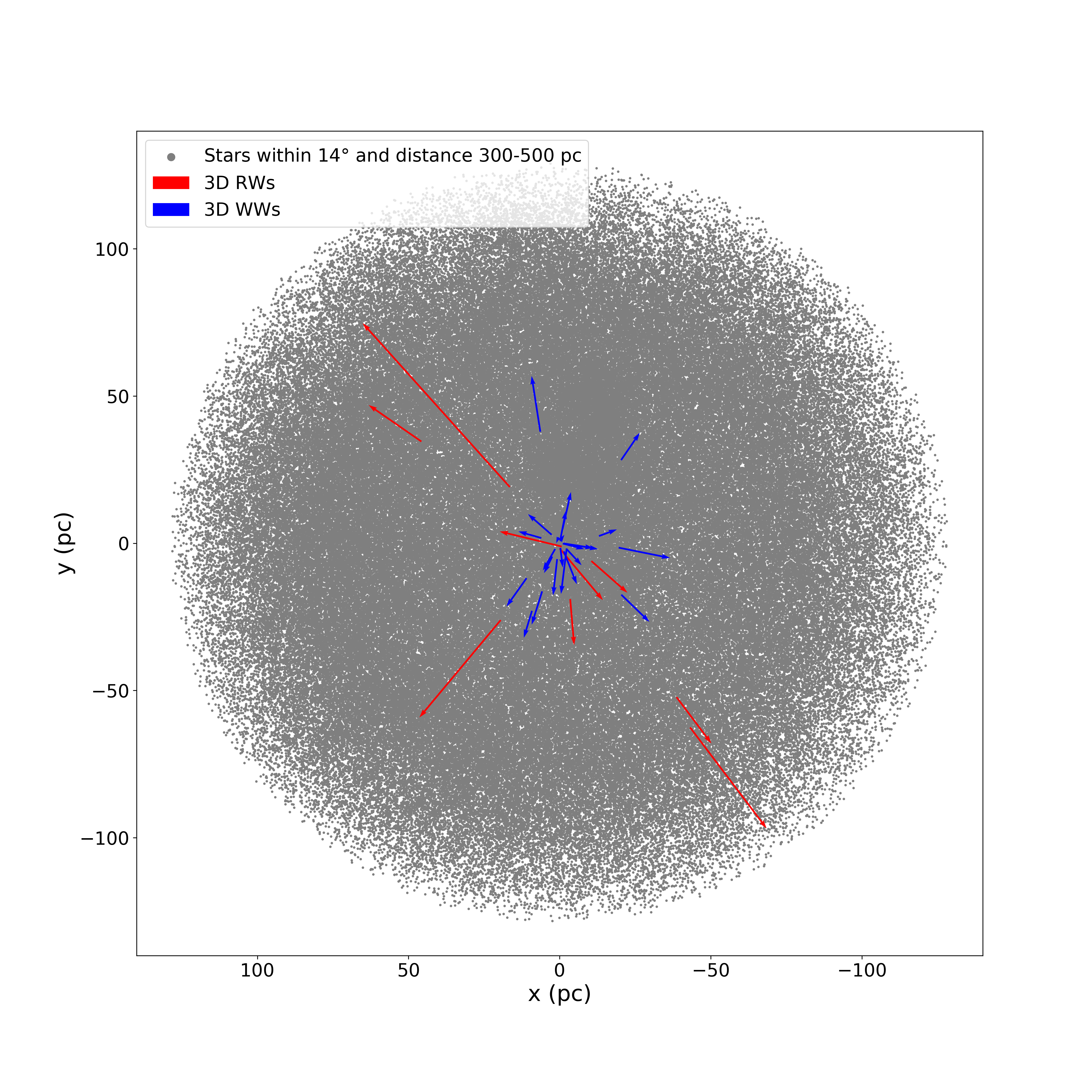}
        \caption{Location of the identified 3D RW and WW stars. The ONC is located in the centre of the plot and extends to a radius of 2.5 pc. We invert the x-axis to replicate the orientation on the sky (i.e. decreasing right ascension from left to right). The 9 RW stars are plotted in ``red'', with the length of the arrows indicating their 2D-velocity in the ONC rest frame, the 24 WW stars are plotted in ``blue''.}
        \label{fig:Cand_location_ONC}
\end{figure*}

\section{Discussion}

The ejected RW stars in our $N$-body simulations quickly leave our 100 pc search region, as most of them have been ejected during the very early dynamical evolution of our simulated star-forming regions. After 2 Myr in the simulations, an average of $\sim$11 (and maximum of 15) RWs are still located within the search area but this average reduces to $\sim$4 (maximum of 10) and $\sim$1 (maximum of 3) RWs after 3 and 4 Myr, respectively. In contrast, all WW stars remain within our 100 pc boundary until at least 3 Myr, only travelling past this boundary towards the end of our simulations.

Our analysis of \textit{Gaia} DR2 finds a total of 31 RW 2D-candidates by tracing back the positions on the sky for up to 4 Myr. Of these RW 2D-candidates, 7 have RV measurements but do not trace back to the ONC in 3D. Nine RW stars can be traced back in three dimensions, all using RVs from \textit{Gaia} DR2.

We find a further potential RW star TYC 4774-868-1 \textit{Gaia} DR2 3209532135777678208), which appears bound when considering only its proper motion but it turns into a RW in three dimensions using its \textit{Gaia} DR2 RV measurement. However, \citet{2015ApJ...807...27C} states a much lower RV of $\sim$28.3 km/s, which results in a RV of $\sim$6.5 km/s in the ONC rest frame instead of 58.5 km/s using \textit{Gaia} DR2 RV. Using this secondary RV source results in a rest-frame space velocity < 10 km/s, which is below our lower boundary for WW stars. Detailed information about this star is shown in Table~\ref{tab:2D_cand_varying_RV} in Appendix~\ref{Varying_RV}.

While simulations show \citep{RN309} that we can expect to find RW stars where only one of their velocity components would identify them as a RW or WW star, there can be other explanations such as the possible binarity of the system. In this case, the difference in RV measurements from two surveys hint at this star being part of a binary, located in the ONC centre.

We also find a potential ejected binary HD 36697 \textit{Gaia} DR2 3016354436766366848) based on the RV measurements by \citet{2015ApJ...807...27C}. This could be either a 3D RW or WW system. Its identification depends on the system's radial velocity, as its 2D-velocity is high enough to put it at least into the WW-velocity regime.

Two of these 9 RWs have mass estimates from the literature putting them all in the low/intermediate-mass category with masses of 1.9--2.5 M$_{\sun}$ \citep{1997AJ....113.1733H,2016ApJ...818...59D}. Based on their position on the CMD, the other 7 3D RW stars are also within the low/intermediate-mass range, as none are consistent with the part of the isochrone that has already reached the MS, i.e. quickly evolving high-mass stars. When we compare the CMD-positions of the 3D RWs to the 3D WWs, we see that they do not extend below the lowest masses of any identified WWs, so we can conclude that none of the RWs identified in our search have a mass < 0.3 M$_{\sun}$. This is a consequence of the magnitude-limit we applied to the CMD, removing any fainter candidates with unreasonably large uncertainties in their astrometric and photometric measurements.

Fig.~\ref{fig:mass_vel_plot} shows that a large number (about half) of all RW and WW stars in our simulations have a mass below 0.3 M$_{\sun}$. This suggests that we can expect to find a large number of RWs/WWs at these low masses. To compare the findings from \textit{Gaia} DR2 to simulations covering the same mass range, we count only low/intermediate-mass RWs within 100 pc with masses between 0.3--8 M$_{\sun}$. We find a RW average of 7.3 $\pm$2.3 at 1 Myr, decreasing to 5.5 $\pm$2.5 at 2 Myr. The number further reduces to 2.2 $\pm$1.4 (0.8 $\pm$0.9) RWs at 3 (4) Myr. The maximum number of RWs from a single simulation reduces to 10 at 2 Myr and 5 (3) RWs at 3 (4) Myr.

Comparing the 9 3D RWs we find in \textit{Gaia} DR2 with our simulation results for the number of RWs gives us an age estimate of $\sim$1.3 Myr when we compare it to the average and $\sim$2.4 Myr when compared to the maximum number from our simulations. This age range is in good agreement with the mean age range (2--3 Myr) of \citet{RN218} and \citet{RN212}. Four of the identified runaways are located within our applied cluster boundary on the sky or are still located within just a few pc of the centre. Several of our RWs only trace back to the region when we consider their large radial velocity and distance errors, making these candidates less certain.

We have not identified any high-mass (> 8 M$_{\sun}$) RW stars in our analysis (neither 2D, nor 3D). We find a B5-star HD 41288 (\textit{Gaia} DR2 3122639449820663040) in 2D, however without an explicit mass estimate from literature. Based on its spectral type, its mass is unlikely to be within our high-mass category. It does not have RV information so cannot be confirmed in 3D.  Two high-mass OB-runaways (AE Aur and $\mu$ Col) have previously been postulated to have originated in the ONC, and \citet{RN325,RN50} suggested these stars were ejected $\sim$2.5 Myr ago. We find a maximum of 2 high-mass RW stars at the end of one of our simulations, however these 2 RWs are not ejected at the same time. The second RW only gets ejected at the very end of this simulation and is in fact a RW-binary. The ejection of high-mass RW stars are rare events in our simulations (as in reality) and in several of our simulations we do not see any high-mass RWs. Due to the IMF, the number of high-mass stars in our simulated regions is low to begin with, which in turn results in only a small number of high-mass RWs, which differs in-between individual simulations.

At lower WW velocities, we find 54 2D-candidates with 30 of them having RV information (either from \textit{Gaia} DR2 or literature sources). We find one O-star $\upsilon$ Ori (\textit{Gaia} DR2 3016424530632449280) WW-candidate in 2D, however its RV measurement \citep{2006AstL...32..759G} prevents it from turning into a 3D-candidate. It is not a known binary \citep{2018A&A...618A.110B}. The Simbad database \citep{2000A&AS..143....9W} lists further RV measurements for this star, none of which is large enough to change this. To become a WW star from the ONC, this O-star would require its RV to point opposite to its current direction. 

We can successfully trace back 26 WW stars to the ONC in 3D, with stellar masses (where available from the literature) between $\sim$0.3--2.7 M$_{\sun}$. Our upper age estimate using 3D RWs is 2.4 Myr and we find 2 3D WWs (\textit{Gaia} DR2 3004263966389331456 and \textit{Gaia} DR2 3217017610938439552) that have flight times since ejection that are larger than this age estimate. These 2 candidates are excluded from the results as they might have come from a different young star-forming region in the neighbourhood of the ONC and we are left with 24 3D WWs.

Within the low/intermediate-mass range > 0.3 M$_{\sun}$, the simulations produce an average of 19.2 $\pm$3.9 WWs within 100 pc at 1 Myr, which increases to 23.1$ \pm$4.2 at 3 Myr. By 4 Myr, we see a reduction in the WW numbers within 100 pc to 21.4 $\pm$4.5, due to stars travelling past this boundary. The maximum number of WWs within 100~pc from a single simulation is 27 WWs at 1 Myr increasing to 32 WWs at 3 Myr then reducing to 30 WWs by 4 Myr. The number of 3D WWs we find in \textit{Gaia} DR2 matches the average number of WWs > 0.3 M$_{\sun}$ in our simulations at $\sim$1.5 Myr. It matches the maximum number found in a single simulation only at an age of $\sim$0.3 Myr, however further identified 3D WWs will increase these age estimates.

Fourteen of the identified 31 RW and 24 of the 54 WW 2D-candidates are missing RV information. From Table~\ref{tab:RWC_2D} and \ref{tab:WWC_2D}, we see that even 2D-candidates with RV are not all confirmed as full 3D-candidates, however 2D-WW candidates with RV measurements appear more often to be 3D WW stars than their RW counterparts. Due to the missing RV information, we are unable to draw any further conclusions from the list of 2D-candidates as RV is required to make an unambiguous RW/WW identification. We have also seen that a high RV might change the RW/WW status of a star that is identified as still bound to the ONC only considering its proper motion.

\textit{Gaia} does not detect any stars in the IR and more than half of the members of the ONC are only detectable in this range \citep{1997AJ....113.1733H}. Fig.~\ref{fig:Cand_location_ONC} shows the location of all identified 3D RW and WWs in relation to the ONC, which is located at the origin in this figure. Several of these are still within the close vicinity (a few pc) of the ONC, we should expect to find further candidates at IR wavelengths. 

Regardless of these limitations, we show with this analysis that the ONC has produced RW and WW stars across the full stellar mass range. In addition to the two known OB-RW stars, we find many more low/intermediate-mass RW and WW stars. This is consistent with the predictions made in the simulations of \citet{RN309}.

A clear 3D-identification is affected by missing or lower-quality radial velocities for many of our 2D-candidates and also by uncertainties in their distances. Furthermore, our analysis is influenced by the uncertainties about the radial extent and distance to the centre of the ONC. Our search region projected on the sky has a diameter of 5 pc, based on the location of existing members. In contrast in the radial direction our search region has a size of 30 pc (15 pc in either direction of the adopted ONC distance of $\sim$400 pc). 

When constructing the CMD, we correct for extinction and reddening, however only a subset of our data has individual $A_{\rm{G}}$ and $E(G_{\rm{BP}}-G_{\rm{RP}})$. This results in us having to estimate values for the remaining stars by averaging over neighbouring stars, leading to highly uncertain age estimates as shown in Table~\ref{tab:RWC_2D} and \ref{tab:WWC_2D}. Even where stars have \textit{Gaia} DR2 measured extinction and reddening values, some have very large errors in these quantities, which can lead to up to a magnitude of error on the CMD. The consequence of these errors are large age ranges for our candidate stars, in particular upper age ranges. \citet{RN303} also note that the \textit{Gaia} DR2 extinction and reddening itself are inaccurate on a star-by-star level. While the age estimates from literature \citep{RN218,2012ApJ...748...14D,2016ApJ...818...59D} are not always consistent with our isochronal age estimates for the confirmed 3D-stars, they still confirm that most of the candidates are younger than the upper age of the ONC. 

The CMDs in Fig.~\ref{fig:2D_RW_candidates_after_correction} and \ref{fig:2D_WW_candidates_after_correction} reveal a large number of 2D trace-backs that are located along the main-sequence and are therefore much older than the ONC's upper age limit. This highlights further that a trace-back on the sky is no indication of the origin of a star without an age estimate. More surprising is the trace-back of several older stars (on the main-sequence with ages of $\sim$5--40 Myr) in 3D, which qualifies these stars as having ``visited'' the ONC in the past. 

It is possible that some of these stars might have even originated in the ONC. \citet{2007ApJ...659L..41P} find lithium depletion in a small group of low-mass stars within the ONC. The authors suggest that this is an indication of these stars being older (>30 Myr) than the rest of the stars in the ONC. However, \citet{2013MNRAS.434..966S} suggest that these differences in lithium are not necessarily evidence of an older population.

\section{Conclusions}

In this paper, we combine \textit{Gaia} DR2 observations with predictions from $N$-body simulations to search for runaway and walkaway stars from the ONC within a distance of 100 pc in an attempt to constrain the region's initial conditions. The conclusions from our simulations and the search in \textit{Gaia} DR2 are summarised as follows:

\renewcommand{\labelenumi}{(\roman{enumi})}
\begin{enumerate}
  \item We find a number of 3D RW (>30\,km\,s$^{-1}$) and 3D WW (10--30\,km\,s$^{-1}$) stars originating from the ONC using \textit{Gaia} DR2 astrometry and photometry in the low/intermediate-mass range (<8 M$_{\sun}$). However, we find no high-mass stars (>8 M$_{\sun}$) in either of the velocity ranges in all three dimensions. About 40 per cent of our 2D-candidates are missing RVs and cannot be confirmed in 3D until this information is available. 
  \item We trace back 9 RWs to the ONC in \textit{Gaia} DR2 that are still within our 100 pc search boundary. Our $N$-body simulations suggest that the older a star-forming region is, the fewer RWs are still found within this boundary. The number of RWs we find in our simulations agrees with those in \textit{Gaia} DR2 when our simulated regions have an age of $\sim$1.3-2.4 Myr. This age estimate for the ONC is in agreement with others from literature \citep{RN218,RN212}.
  \item Our simulations predict that all WWs are still to be found within the search region until at least 3 Myr and that the number of WWs increases up to this age. In \textit{Gaia} DR2, we find 26 WWs all with masses between 0.3 and 2.7 M$_{\sun}$. Twenty-four of those have been ejected within the past 2.4 Myr (upper age implied from our RW findings). This agrees with the average number of WWs at an age of $\sim$1.5 Myr from our simulations, but is below the maximum from a single simulation at virtually any age. However, future \textit{Gaia} data releases and complementary IR surveys may enable us to identify further WWs, which will increase these age estimates.
  \item Our analysis shows that ejected stars might be useful in constraining the initial conditions of star-forming regions. However to use this method to its full extent requires further improvements in \textit{Gaia} or other observations, e.g. more measurements of radial velocities in addition to proper motion; extinction and reddening values for a larger number of stars.
\end{enumerate}

\section*{Acknowledgements}

We thank Anthony Brown for his help related to using the \textit{Gaia} DR2 dataset, in particular RUWE-filtering and Michael Kuhn for his help with the analysis process and for providing a list of source IDs for the ONC.   

CS acknowledges PhD funding from the 4IR STFC Centre for Doctoral Training in Data Intensive Science. CS thanks ESTEC/ESA in Noordwijk for hosting her as a visiting researcher during the major part of the work on this project and the \textit{Gaia} team there for their support and feedback. RJP acknowledges support from the Royal Society in the form of a Dorothy Hodgkin Fellowship.

This work has made use of data from the European Space Agency (ESA) mission {\it Gaia} (\url{https://www.cosmos.esa.int/gaia}), processed by the {\it Gaia} Data Processing and Analysis Consortium (DPAC, \url{https://www.cosmos.esa.int/web/gaia/dpac/consortium}). Funding for the DPAC has been provided by national institutions, in particular the institutions participating in the {\it Gaia} Multilateral Agreement. 

This research has made use of the SIMBAD database, operated at CDS and the VizieR catalogue access tool, CDS, Strasbourg, France.

%%%%%%%%%%%%%%%%%%%%%%%%%%%%%%%%%%%%%%%%%%%%%%%%%%



\bibliographystyle{mnras}
\bibliography{Main_document} % if your bibtex file is called example.bib

%%%%%%%%%%%%%%%%%%%%%%%%%%%%%%%%%%%%%%%%%%%%%%%%%%

%%%%%%%%%%%%%%%%% APPENDICES %%%%%%%%%%%%%%%%%%%%%

\appendix

\section{Excluded candidates}\label{App_excluded_cand}

Table~\ref{tab:RWC_2D_excluded} provides information on the identified 2D-candidates that have been excluded, either due to their RV pointing towards the ONC or because their required RV due to their radial position would be above |500|\,km\,s$^{-1}$. 

\begin{table*}
%	\centering
	\caption{Excluded RW and WW star 2D-candidates. Column 2+3: velocity in ONC rest frame [rf]; Column 3: RV sources -- $^{a}$\textit{Gaia} DR2, $^{b}$\citet{2015ApJ...807...27C},  $^{c}$\citet{2006AstL...32..759G}; Column 4: Reason for exclusion;  Column 5--7: from literature sources -- $^{1}$\citet{2016ApJ...818...59D},   $^{2}$\citet{1999MSS...C05....0H}, $^{3}$\citet{1995A&AS..110..367N}, $^{4}$\citet{1988mcts.book.....H}, $^{5}$\citet{RN263}.}
	\label{tab:RWC_2D_excluded}
	\begin{tabular}{lcccccc} % four columns, alignment for each
		\hline
		\textit{Gaia} DR2 source-id & 2D-velocity rf & Radial velocity rf & Exclusion reason & Age & Mass & Spectral type \\
		& (km\,s$^{-1}$) & (km\,s$^{-1}$) & & (Myr) & (M$_{\sun}$) &\\
		\hline
		Excluded RW candidates\\
		\hline
        3011539434830106624	&64.5 $\pm$0.4&	-26.5 $\pm$6.8$^{a}$ & RV points towards ONC& - & - & -\\
        3016948241764201472 &50.2 $\pm$0.8&-	&RV required > |500|\,km\,s$^{-1}$& - & - & -\\
        3187254518368160000	&47.5 $\pm$0.4&	-66.3 $\pm$6.6$^{a}$ & RV points towards ONC & - & - & -\\
        3017157630011568000 &42.8 $\pm$0.4    & -57.6$\pm$6.6$^{b}$   & RV points towards ONC & 131$^{1}$& 1.4$^{1}$ & - \\
        3011184292574204672	&41.7 $\pm$0.4   &	15.0 $\pm$6.6$^{a}$ & RV points towards ONC & -& - & -  \\
        3017247618166246784	&39.4 $\pm$0.4&	86.1 $\pm$6.6$^{a}$ &  RV points towards ONC& 2.0$^{1}$ & 2.2$^{1}$& - \\
        3206712880587397120 &38.9 $\pm$0.4& -16.7 $\pm$6.6$^{a}$ & RV points towards ONC & - & - & -\\
        3122079111207514496 &32.7 $\pm$0.4& 28.2 $\pm$6.6$^{a}$  & RV points towards ONC & - & - & -\\
		\hline
	    Excluded WW star candidates\\
		\hline
		3209554744485606400	&24.5 $\pm$0.7& - & RV required > |500|\,km\,s$^{-1}$& - & - & -\\
    	3013314424554785024 &23.9 $\pm$0.4& -17.2 $\pm$6.6$^{a}$ & RV points towards ONC & - & - & G8/K0$^{2}$ \\
    	2989308443587969664	&18.5 $\pm$0.3	&   0.3	$\pm$6.6$^{a}$ & RV points towards ONC & - & - & F6/7$^{4}$ \\
        3017240746218498176	&17.9 $\pm$0.6 & -2.3 $\pm$6.6$^{b}$ & RV points towards ONC & - & - & - \\
        3015045708692252672 &17.4 $\pm$0.5& -11.4 $\pm$6.6$^{a}$ & RV points towards ONC & - & - & - \\
        3023538370864242688	&17.3 $\pm$0.5	&-19.2$\pm$6.6$^{b}$ & RV points towards ONC & - & - &A7$^{3}$ \\
        3019516254250648832 &14.6 $\pm$0.4& -23.2 $\pm$6.7$^{a}$&RV points towards ONC& - & - & - \\
        3215185309169853568	&11.1 $\pm$0.6 & 8.1 $\pm$6.9$^{c}$&RV points towards ONC& 0.2$^{5}$ & 7.9$^{5}$ & B3$^{2}$ \\
        3017244216552060672	&0.5 $\pm$0.4 &15.6 $\pm$9.8$^{a}$& RV points towards ONC & 5.7$^{1}$ & 1.3$^{1}$ & -\\ 
		\hline
	\end{tabular}
\end{table*}

\section{2D-candidates with multiple, varying RV measurements}\label{Varying_RV}

Table~\ref{tab:2D_cand_varying_RV} provides information on identified 2D-candidates that would not be identified as RWs or WWs only by their 2D-velocity, but could be when considering their RV measurements. We have identified several such stars in Table~\ref{tab:WWC_2D}, but also have stars with multiple RV measurements between different literature sources, which could also be indicating a binary system.

\begin{table*}
%	\centering
    \renewcommand\arraystretch{1.05}
	\caption{Slow 2D-candidates with multiple, varying RV measurements; Column 2: velocity in ONC rest frame [rf]; Column 3: RV sources of varying measurements - $^{a}$\textit{Gaia} DR2, $^{b}$\citet{2015ApJ...807...27C}, $^{c}$\citet{2018AJ....156...84K}; Column 4: indication of 3D-candidate status; Column 5: minimum flight time since ejection (crossing of search boundary); Column 6: age from PARSEC isochrones \citep{RN225}; Column 7--9: from literature sources - $^{1}$\citet{1997AJ....113.1733H}, $^{2}$\citet{2016ApJ...818...59D}.}
	\label{tab:2D_cand_varying_RV}
	\begin{tabular}{lcccccccc} % four columns, alignment for each
		\hline
		\textit{Gaia} DR2 source-id & 2D-velocity rf & Radial velocity rf & 3D-cand. & Flight time & Iso. age & Age & Mass & Spectral type  \\
		 & (km\,s$^{-1}$) & (km\,s$^{-1}$) & & (Myr) & (Myr) & (Myr) & (M$_{\sun}$) & \\
		\hline
		3209532135777678208	&3.2	$\pm$0.4	&  a, b	&	?	&	in cluster	& 5.0\,$^{+5.0	}_{	-3.0}$ & 5.6$^{2}$ & 2.1$^{2}$ & -  \\
		3017270669252519680	&	2.9	$\pm$0.4	&	a, b	&	?	&	in cluster		& 1.0\,$^{+	20.0	}_{	-0.9	}$ &	0.6$^{2}$	&	1.0$^{2}$	&-		\\
        3209528012609165440	&	2.5	$\pm$0.3	&	a, b	&	?	&	in cluster		&2.5\,$^{+	4.7	}_{	-2.2	}$ &	1.4$^{2}$ &	0.9$^{2}$	&	-	\\
        3209521037582290304	&	1.7	$\pm$0.4	&  a, c	& ?	&	in cluster		& 0.3	$\pm$0.2&	0.3$^{2}$	&	1.6$^{2}$	&K3$^{1}$	\\
		\hline
	\end{tabular}
\end{table*}

\newpage
\section{Past visitors to the ONC}\label{App_visitors}

Table~\ref{tab:Visitors_list} provides information on sources that can be successfully traced back in 3D to the ONC in the past 2.4 Myr (upper age limit implied by the identified RW stars) but their position on the CMD identifies them as MS stars or older pre-MS stars.

\begin{table*}
%	\centering}
	\caption{Past visitors to the ONC. Column 2+3: velocity in ONC rest frame [rf]; Column 3: RV sources -- $^{a}$\textit{Gaia} DR2; Column 4: age from PARSEC isochrones \citep{RN225}; Column 5--7: from literature sources -- $^{1}$\citet{2016ApJ...818...59D}.}
	\label{tab:Visitors_list}
	\begin{tabular}{lccccccc} % four columns, alignment for each
		\hline
		\textit{Gaia} DR2 source-id & 2D-velocity rf & Radial velocity rf & Iso. age & Age & Mass & Spectral type \\
		 & (km\,s$^{-1}$) & (km\,s$^{-1}$) & (Myr) & (Myr) & (M$_{\sun}$) & \\
		\hline
        Visitors at RW velocities\\		
		\hline
        2986529565387616896	&	85.1	$\pm$0.3	&	-18.3	$\pm$6.8$^{a}$& $\sim$22--30 & - & - & - \\
        2976576167656119296	&   80.9	$\pm$0.3    &	-35.1   $\pm$6.7$^{a}$&  $\sim$30 & - & - & -\\
        3240725452454418048	&	64.4	$\pm$0.4	&	14.0	$\pm$6.7$^{a}$& $\sim$18--20 & - & - & -\\
        3014832609594260224	&   63.1	$\pm$0.4    &	-16.0   $\pm$19.5$^{a}$&$\sim$15--20 & - & - & -\\
        3009308457018637824	&	58.1	$\pm$0.3	&	-66.2	$\pm$6.7$^{a}$& $\sim$25--30 & - & - & -\\
        3021115184676332288	&	57.8	$\pm$0.4	&	10.1    $\pm$6.6$^{a}$& $\sim$20 & - & - & -\\
        2992509671692197632 &   57.2    $\pm$0.3	&	-36.7	$\pm$6.7$^{a}$& $\sim$25 & - & - & -\\
        2981722809790374400	&   52.8	$\pm$0.3    &	-45.5	$\pm$6.6$^{a}$& $\sim$25--30 & - & - & -\\
        2989584932106141184	&   51.6	$\pm$0.3    &	-22.9	$\pm$6.6$^{a}$& $\sim$20--25 & - & - & -\\
        3024722888484450944	&	51.6	$\pm$0.4	&	23.7	$\pm$6.7$^{a}$& $\sim$17--20 & - & - & -\\
        2971498824821941760	&	46.6	$\pm$0.4	&	-26.1	$\pm$6.7$^{a}$& $\sim$17--20 & - & - & -\\
        3240501873637049856	&   46.5	$\pm$0.4	&   -16.7	$\pm$6.7$^{a}$& $\sim$20 & - & - & -\\
        2998151094058000128	&	44.5	$\pm$0.4	&	-34.3	$\pm$6.7$^{a}$& $\sim$28--30 & - & - & -\\
        3015908138125994112	&	42.2	$\pm$0.4	&	28.9	$\pm$6.8$^{a}$& $\sim$20 & - & - & -\\
        3005776138475075712	&   38.5	$\pm$0.4	&   10.6	$\pm$9.2$^{a}$& $\sim$30--35& - & - & -\\
        3014762309569718272	&   36.3	$\pm$0.3	&   -45.8	$\pm$6.7$^{a}$& $\sim$30 & - & - & -\\
        3228935835946246272	&   33.8	$\pm$0.4	&   3.1 	$\pm$6.7$^{a}$& $\sim$12--15 & - & - & -\\    
        2989747213149681408	&	32.4	$\pm$0.4	&	-4.7	$\pm$6.7$^{a}$& $\sim$25--30 & - & - & -\\
        3017817268267372032	&	30.2	$\pm$0.4	&	-25.5	$\pm$7.7$^{a}$& $\sim$30--35 & - & - & -\\
        3011733017597107840	&	29.9	$\pm$0.4	&	-14.0	$\pm$6.8$^{a}$& $\sim$20--22 & - & - & -\\
        3010392369324807808 &   27.5	$\pm$0.4    &	-12.9   $\pm$6.8$^{a}$& $\sim$30 & - & - & -\\
        3207687464502067456	&	27.4	$\pm$0.5	&   34.2	$\pm$7.7$^{a}$& $\sim$12--15 & - & - & -\\
        3022827296078444672	&	25.6	$\pm$0.4	&	29.4	$\pm$6.7$^{a}$& $\sim$17--20 & - & - & -\\
        3019025799048789504	&	22.7	$\pm$0.4	&	22.8	$\pm$6.6$^{a}$& $\sim$17--20 & - & - & -\\        
        3017348803299280896 &	22.6	$\pm$0.4	&   -46.1	$\pm$6.7$^{a}$& $\sim$25--30 & 27.0$^{1}$ & 1.1$^{1}$& -\\
        2999242806023093888	&   21.4	$\pm$0.3	&	-53.0	$\pm$6.7$^{a}$& $\sim$30 & - & - & -\\
        3210090515884826752	&   21.1	$\pm$0.4	&	23.7	$\pm$7.0$^{a}$& $\sim$25--30 & - & - & -\\
        3209536636903447936	&   15.4	$\pm$0.4	&	-38.8	$\pm$6.7$^{a}$& $\sim$12--17 & 16.2$^{1}$ & 1.4$^{1}$& -\\
        3011879768037861504 &   13.3	$\pm$0.4    &	-28.7   $\pm$6.6$^{a}$& $\sim$10--15 & - & - & -\\
		\hline
	    Visitors at WW velocities\\
		\hline
        2996472071080530176	&	24.5	$\pm$0.3	&	-14.5	$\pm$6.9$^{a}$& $\sim$30 & - &- & -\\
        3016578221743133952	&   23.8	$\pm$0.3	&   8.7 	$\pm$6.8$^{a}$& $\sim$35--40  & - & - & -\\
        3219402241203000576	&   23.2	$\pm$0.4    &	-0.8	$\pm$7.2$^{a}$& $\sim$30  & - & - & -\\
        3010331032894313984	&   22.5	$\pm$0.4    &	-8.3	$\pm$6.7$^{a}$& $\sim$7--10 & - & - & -\\
        3010652777484841344	&	20.6	$\pm$0.4	&	-1.6	$\pm$6.7$^{a}$& $\sim$28--30 & - & - & -\\
        3010517434474967040	&   19.9	$\pm$0.3	&	6.1	    $\pm$12.2$^{a}$& $\sim$5--20  & - & - & -\\
        3208716023269799808	&	19.8	$\pm$0.4	&	0.3	    $\pm$13.7$^{a}$& $\sim$30--32 & - & - & -\\
        3009639233922459008	&   19.3	$\pm$0.3	&	-20.8	$\pm$7.6$^{a}$& $\sim$30 & - & - & -\\
        3207885750255374336 &   18.3	$\pm$0.3    &	-17.5   $\pm$7.4$^{a}$& $\sim$40 & - & - & -\\
        3022823615290307840	&	17.1	$\pm$0.4	&	-23.4	$\pm$6.7$^{a}$& $\sim$20 & - & - & -\\
        3014826802798497920	&	17.1	$\pm$0.4	&	-8.4	$\pm$7.5$^{a}$& $\sim$30--32 & - & - & -\\
        3012237036301375872	&   16.8	$\pm$0.4	&	-13.5	$\pm$7.6$^{a}$& $\sim$20--25 & 46.0$^{1}$ & 1.2$^{1}$& -\\
		3210977649969156224	&   13.7	$\pm$0.3	&	1.1    $\pm$6.6$^{a}$& $\sim$20--35 & - & - & -\\
		\hline
	\end{tabular}
\end{table*}

%%%%%%%%%%%%%%%%%%%%%%%%%%%%%%%%%%%%%%%%%%%%%%%%%%

% Don't change these lines
\bsp	% typesetting comment
\label{lastpage}
\end{document}